\documentclass[aps,prl,twocolumn,amsmath,amssymb,showpacs,superscriptaddress]{revtex4-2}
\usepackage[colorlinks=true,citecolor=blue,urlcolor=blue]{hyperref}
\usepackage{graphicx}
\usepackage{dcolumn}
\usepackage{bm}
\usepackage{tabularx}
\usepackage[margin=0.6in]{geometry}
\usepackage{xr}
\usepackage[braket, qm]{qcircuit}
\DeclareMathOperator{\Tr}{Tr}
\usepackage{xr}
\usepackage{svg}
\usepackage{cleveref}
\usepackage{arabtex}
\usepackage{url}

\makeatletter

\newcommand*{\addFileDependency}[1]{argument=file name and extension
\typeout{(#1)}
\@addtofilelist{#1}
\IfFileExists{#1}{}{\typeout{No file #1.}}
}
\makeatother

\begin{document}

\title{Single-Layer Digitized-Counterdiabatic Quantum Optimization for {\it p}-spin Models}

\author{Huijie Guan}
\thanks{These two authors contributed equally}
\email{a0610328guanguan@gmail.com}
\affiliation{QuantumCtek Co.,Ltd. Science Park, No.777 Huatuo Lane, High-tech Industrial Development Zone, Hefei, Anhui, China}

\author{Fei Zhou}
\thanks{These two authors contributed equally}
\email{zhoufei@jiqt.org}
\affiliation{Jinan Institute of Quantum Technology, Lixia District747 Shunhua Rd, Ji'nan, Shandong, China}

\author{Francisco Albarr\'an-Arriagada}
\email{francisco.albarran@usach.cl}
\affiliation{Departamento de F\'isica, Universidad de Santiago de Chile (USACH), Avenida V\'ictor Jara 3493, 9170124, Santiago, Chile}
\affiliation{Center for the Development of Nanoscience and Nanotechnology 9170124, Estaci\'on Central, Santiago, Chile}

\author{Xi Chen}
\email{xi.chen@ehu.eus}
\affiliation{Department of Physical Chemistry, University of the Basque Country UPV/EHU, Apartado 644, 48080 Bilbao, Spain}
\affiliation{EHU Quantum Center, University of the Basque Country UPV/EHU, Barrio Sarriena, s/n, 48940 Leioa, Spain}

\author{Enrique Solano}
\email{enr.solano@gmail.com}
\affiliation{Kipu Quantum, Greifswalderstrasse 226, 10405 Berlin, Germany}

\author{Narendra N. Hegade}
\email{narendrahegade5@gmail.com}
\affiliation{Kipu Quantum, Greifswalderstrasse 226, 10405 Berlin, Germany}
\thanks{These three authors contributed equally}

\author{He-Liang Huang}
\email{quanhhl@ustc.edu.cn}
\affiliation{Henan Key Laboratory of Quantum Information and Cryptography, Zhengzhou, Henan 450000, China}

\maketitle

\textbf{Quantum computing holds the potential for quantum advantage in optimization problems, which requires advances in quantum algorithms and hardware specifications. Adiabatic quantum optimization is conceptually a valid solution that suffers from limited hardware coherence times. In this sense, counterdiabatic quantum protocols provide a shortcut to this process, steering the system along its ground state with fast-changing Hamiltonian. In this work, we take full advantage of a digitized-counterdiabatic quantum optimization (DCQO) algorithm to find an optimal solution of the $p$-spin model up to 4-local interactions. We choose a suitable scheduling function and initial Hamiltonian such that a single-layer quantum circuit suffices to produce a good ground-state overlap. By further optimizing parameters using variational methods, we solve with unit accuracy 2-spin, 3-spin, and 4-spin problems for 100\%, 93\%, and 83\% of instances, respectively. As a particular case of the latter, we also solve factorization problems involving 5, 9, and 12 qubits. Due to the low computational overhead, our compact approach may become a valuable tool towards quantum advantage in the NISQ era.}

\section{Introduction}

Quantum Computing might solve optimization problems that are intractable with classical algorithms~\cite{Huang2023NearTerm}. Usually, optimization methods are formulated as Ising spin-glass models whose ground states encode the solutions of the original problem. In principle, the ground state of a given problem Hamiltonian can be obtained by adiabatically evolving from the ground state of a simpler one, say a local transverse-field case. However, adiabatic quantum computing requires operation times exceeding the hardware coherence times, a problem that gets worse as the number of qubits increases~\cite{yamashiro2019dynamics}. 

On the other hand, adding counterdiabatic (CD) terms allow for a controlled acceleration of the adiabatic quantum dynamics~\cite{guery2019shortcuts}, borrowing from the frame of shortcuts to adiabaticity~\cite{demirplak2003adiabatic, demirplak2005assisted, berry2009transitionless,Chenprl2010,del2012assisted}. Unfortunately, the exact calculation of the CD terms requires the knowledge of spectral properties~\cite{kolodrubetz2017geometry}. Moreover, its experimental implementation would involve nonlocal terms, compromising the limited connectivity of current quantum processors. In this regard, approximated CD terms can be obtained variationally from a user-defined ansatz which minimizes transitions among eigenstates as much as possible. This approach complies with existing constraints like locality~\cite{sels2017minimizing}, requiring less overhead from gate decomposition and transpilation.

In order to profit from the hardware advances in digital quantum computing, we can digitize the approximate counterdiabatic dynamics, making use of the purely-quantum and hybrid classical-quantum DCQO paradigms \cite{hegade2022digitized, cadavid2023efficient}. In this work, we take advantage of them to study the ground states of the $p$-spin model with up to 4-local interactions. The latter case of $p=4$ provides a solution for the factorization problem~\cite{hegade2021digitized,anschuetz2019variational}, $p=2$ solves the portfolio optimization problem~\cite{hegade2022portfolio}, or we may even consider randomly generated cases as in mixed $p$-spin glass models~\cite{dey2022hypergraph}. As a key improvement, we prove that a single-layer hybrid DCQO with warm-started initialization of the CD ansatz produces unit accuracy for $p$-spin models with $p=2,3$, and $4$, including prime factorization in the latter.

The $p$-spin model under examination has a wide array of applications, which have been thoroughly explored in various studies, such as the fully connected pure $p$-spin model \cite{gross1984simplest, castellani2005spin}, the p-XORSAT model~\cite{mezard2003two}, K-SAT~\cite{mezard2009information}, q-coloring~\cite{mezard2009information}, and numerous others when $p(q) \leq 4$. Additionally, for the case of $p=2$, the Hamiltonian of the system includes well-known models that involve pairwise interactions, like MaxCut [13], the Sherrington-Kirkpatrick (SK) model [14], the Maximal Independent Set [4], as well as other quadratic unconstrained optimization problems. 

The manuscript is arranged as follows. In Sec. II, we formulate the DCQO approach. In Sec. III, we present the results of the factorization problem. In Sec. IV, we apply the approach to mixed $p$-spin problems. Finally, in Sec. V, we discuss our results and conclude.

\section{Methodology}
In order to prepare the ground state of a targeted Hamiltonian, the system is engineered to evolve under the following dynamics
\begin{align}\label{eq:h_ad}
\mathcal{H}(t) = (1- \lambda(t))H_i + \lambda(t) H_f .
\end{align}
Here, $H_i = -h_x \sum_i \sigma_x^i$ is a transverse-field Hamiltonian and the final one corresponds to the problem of interest. $\lambda(t)$ is a user-defined scheduling function, that drives $\mathcal{H}$ from $H_i$ to $H_f$. The scheduling function is set as $\lambda(t) = \sin^2\left(\pi/2 \sin^2\left(\frac{\pi t}{2\tau}\right)\right)$, whose derivative vanishes for $t = 0$ and a given $\tau$~\cite{sels2017minimizing}. The initial state is prepared as $|-\rangle^{\otimes n}$, which is the ground state for $\mathcal{H}(t = 0)$. 

For a time-dependent Hamiltonian, its eigenstate would vary with time as well. For a moving observer in the instantaneous eigenbasis of $\mathcal{H}$, the Hamiltonian picks up an extra contribution and becomes
\begin{align}
\mathcal{H}^{\text{eff}} = \tilde{\mathcal{H}} - \dot{\lambda}\tilde{A} ,
\end{align}
where $\dot{\lambda} = d\lambda/dt$ is changing rate and $\tilde{\mathcal{H}}$ is the Hamiltonian and $\tilde{A}$ is adiabatic gauge potential in the moving frame which is the origin of excitations. The counterdiabatic approach works by adding a CD term that exactly cancels this gauge potential, i.e.
$H_{CD} = \mathcal{H} + \dot{\lambda}A$. 

The matrix of $A$ can be written down as the expectation value of $\partial_\lambda \mathcal{H}$ (supplementary material), but its explicit form is very hard to solve. A variational approach has been proposed to find optimal CD terms within a user-defined ansatz \cite{sels2017minimizing}. The variational procedure seeks the solution that minimizes action $\mathcal{S}$ defined by
\begin{align}\label{eq:3}
G(A) = \partial_\lambda \mathcal{H} + i[A(\lambda), \mathcal{H}] , \notag \\
\mathcal{S}(A(\lambda) = \Tr[G^2(A(\lambda)] \, , \,
\frac{\delta S(A(\lambda))}{\delta A(\lambda)} = 0 \, .
\end{align}
Here, $A$ is approximated CD terms with variable parameters. One way to construct $A$ is to use the nested commutator formula expressed as a sum of different orders of nested commutator between $\mathcal{H}$ and $\partial_{\lambda}\mathcal{H}$ \cite{claeys2019floquet}. Although the expression can be proved exact with infinite order, the truncated result cannot be properly justified. In the supplementary material, we have made a comparison between first and second-order results, where we confirm that the second order is indeed large for a particular problem we will consider.

Another way to build $A$ is by user-defined ansatz from Pauli operators. Among $4^n -1$ Pauli operators, only a subset will contribute. From the matrix form of $A$, one can tell that $A$ has to be imaginary. Thus $A$ should be composed of terms with an odd number of Pauli-$y$ operator. Below, we will consider various combinations of $Y$ and $YZ$ types of ansatz, see the next section for more details. Although $YZZ$, $XYZ$ type of terms are also relevant for the exact CD ansatz, we will restrain to 2-local interactions to simplify both the theory as well as implementation.

Given the CD ansatz, the final state can be obtained as
\begin{align}
|\psi(\tau)\rangle = U(t)|\psi(0)\rangle ,
\end{align}
where
\begin{align}\label{eq:14}
&U(\tau) = \mathcal{T} \exp(-i\int dt H(t)) \notag\\
&\approx \prod_{j = 0}^{\text{t}\text{-step}} \exp(-i \Delta t \mathcal{H}(t_j)) \exp(-i \Delta \lambda_j \mathcal{A}(\lambda_j)) .
\end{align}
Here, $t_j = j\Delta t$, $\lambda_j = \lambda(t_j)$, and $\Delta t$ is a small time step. Trotterization is employed to split exponentiation of integration into (t-step + 1)  products. The replacement of $\Delta t$ by $\Delta \lambda$ in the second term is due to the factor $\dot{\lambda}$ in $H_{CD}$. Note that the first term would vanish with diminishing $\Delta t$, while the second term would still survive. Therefore, in the quenching limit when $\tau\to0$, the contribution from the $\mathcal{H}$ can be neglected for $t \neq 0\text{ or }\tau$.

As shown in the Supplementary Material, the strength of each CD term forms a bell shape which depends on $h_x$. By tuning $h_x$, it is possible to make each term peak around $t = \tau/2$. Moreover, with the choice of the scheduling function $\lambda(t)$, $\dot{\lambda}$ also reaches its maximum at the middle point. This suggests the feasibility of taking t-step = 2. In this setup, Equation~\eqref{eq:14} becomes
\begin{align}
U(\tau) = \exp\left(-i\Delta\lambda A(\lambda(\tau/2)\right) ,
\end{align}
where $\Delta\lambda = \dot{\lambda}(\tau/2)\Delta t$. Note that at $t = 0 $, the CD contribution vanishes due to zero changing rate and the system state is an eigenstate of the $H_0$, adding only a pure phase. Similarly, at $t = \tau$, the CD term disappears. In the ideal case, the final state is an eigenstate of $H_f$, providing another phase. This results in a single-layer construction of the DCQO algorithm.

We recall that many approximations have been made so far, which prevent the algorithms from steering the system states along the ground state of the instantaneous Hamiltonian. For example, the 2-local CD ansatz is only approximate, so the decision to stop the evolution operator at $\tau$ is rough and trotterization brings more errors. It should be no surprise that the final state is not a pure ground state, but a distribution over a different configuration basis. However, the variational approach seeks the best solutions to reduce the distance between the true ground state and the approximated one. Therefore, as we will show, using such a solution as an initialization will in most cases warm-start the optimization process and avoid local minimum as encountered in many variational quantum algorithms.

\section{Factorization Problem}
The factorization problem can be cast into an Ising spin-glass model by encoding the binary representation of the factors as well as carry bits in the spin degrees of freedom. The multiplication equation produces a set of bit-wise constraint equations, whose violation would lead to an energy penalty in the cost function, similar to the treatment of a K-SAT problem. 

To further reduce the source requirement, classical preprocessing is implemented to reduce the number of constraint equations based on the binary properties of each bit, e.g. $p^2 = p$, $pq = 1\Rightarrow p = q = 1$, etc. Following the lead in Ref.~\cite{anschuetz2019variational}, we introduce two improvements. Firstly, we modify the simplification rules for the preprocessing codes. The change solves the randomness problem of the original algorithms and reduces some cases to trivial ones completely solvable with classical preprocessing. Removing superficial complexity is of great importance for the demonstration of quantum factoring algorithms. The Ising spin problem may have a decoupled set of spins if constraint equations are not fully simplified. In the meanwhile, special care has been made to ensure that the preprocessing rules do not oversimplify the calculation as well~\cite{smolin2013oversimplifying}. Here, the complexity of the preprocessing algorithm is $O(n^3)$~\cite{anschuetz2019variational}. See Supplementary Material for more details.

As an example, we study the factorization problem of 1261, 767, and 9983 which reduce to 5, 9, and 12 qubit Ising model after classical preprocessing. The explicit form of the Ising model is listed in Supplementary Material. To conserve space, we will denote these models as $H^{5q}$, $H^{9q}$ and $H^{12q}$. $H^{5q}$ is a mixed 3-spin problem, while $H^{9q}$ and $H^{12q}$ are mixed 4-spin problems. In all of the cases, the high-order interactions are sparse. 

\begin{figure}
\centering
\includegraphics[page = 1, width = 0.45\textwidth]{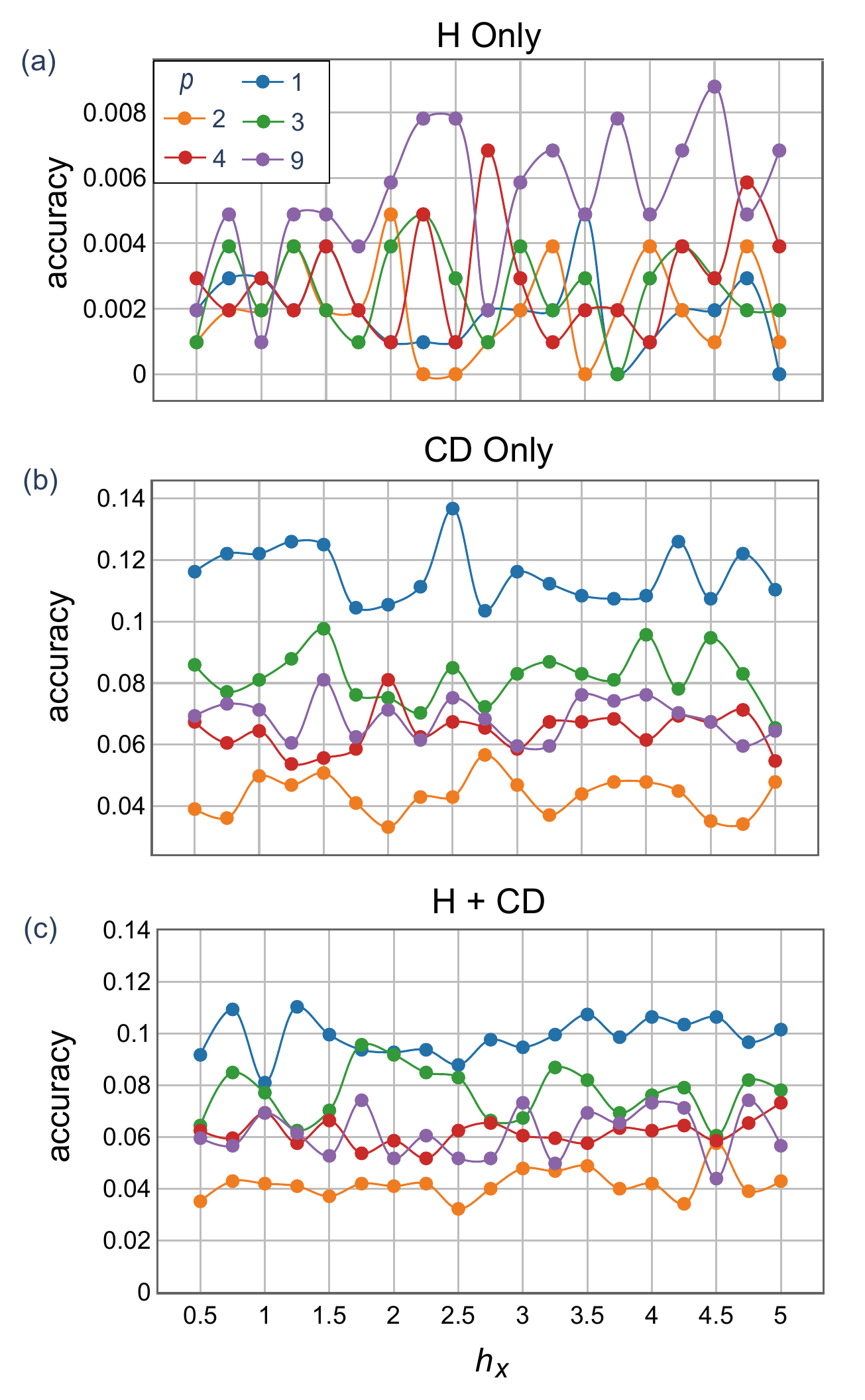}
\caption{Accuracy of algorithms for solving factorization problem $H^{9q}$ with different evolution schemes $U(t)$. From top to bottom, the algorithm considers evolution with problem Hamiltonian only(a), counterdiabatic terms only (b), and both problem Hamiltonian and counterdiabatic terms(c) respectively. $x$ axis denotes initial Hamiltonian $h_x$. Each line corresponds to a different layer number $p$. For counterdiabatic term, the ansatz $A = \sum_i \alpha_i J_i\sigma_y^i + \sum_{i<j} (\beta_{ij}J_{ij}\sigma_y^i\sigma_z^j + \gamma_{ij}J_{ij}\sigma_z^i \sigma_y^j)$ is considered. The time step $\Delta t$ is set to 0.1.}
\label{fig1}
\end{figure}

Applying the CD approach outlined in Section II, we obtain the final state distribution for $H^{5q}$, $H^{9q}$ and $H^{12q}$ with CD ansatz chosen as $A = \sum_i \alpha_i J_i\sigma_y^i + \sum_{i<j} (\beta_{ij}J_{ij}\sigma_y^i\sigma_z^j + \gamma_{ij}J_{ij}\sigma_z^i \sigma_y^j)$. Fig.~\ref{fig1} displays the performance of algorithms in finding the ground state of $H^{9q}$ with different evolution operators $U(t)$, initial Hamiltonians $h_x$, and the number of trotter steps. Here we adopt parameter $p = n_{\text{step}}- 1$ as layer number which is more commonly used in QAOA studies. Different plots relate to different evolution schemes. In each plot,  $p = 1,2,3,4,9$ are studied and plotted with different colored lines. The x-axis denotes $h_x$. For all situations, the lines wiggle in the $x$ direction, indicating complex dependence of performances to the initial Hamiltonian.  In Fig.~\ref{fig1}(a), the evolution is carried out with the problem Hamiltonian only, displaying an accuracy of around 1\% for various time steps and initial Hamiltonian $h_x$.  We observed slightly better performance with larger time steps, which relates to a longer evolution time. Lines in   \ref{fig1}(b) describe evolution with CD terms only,  neglecting contributions from the problem Hamiltonian.

The algorithm produces an accuracy of around 10\%, which is a great improvement over the bare evolution with $\mathcal{H}$. This is because in the fast-changing scenario, the gauge potential $\dot{\lambda}\mathcal{A}$ has a more prominent contribution compared to the original Hamiltonian, thus compensating for this term would greatly suppress excitations while ignoring the Hamiltonian will have less effect. We also observe that the accuracy of algorithms with odd $p$ have better performance than those of even ones. This is due to the fact that the counterdiabatic contribution from $t = \tau/2$ is in general greatest, which is only sampled for odd $p$. Interestingly, the one with $p =1$ has the best performance in general. This is partly because larger $p$ indicates longer evolution time and weaker CD terms. We also compare the performance with algorithms considering both CD contributions and problem Hamiltonian, the results are shown in Figure \ref{fig1}(c). To our surprise, having full counterdiabatic Hamiltonian in the evolution leads to a slight decrease in accuracy for all cases considered, especially for $p = 1$. This indicates that keeping only counterdiabatic terms in the evolution not only simplifies implementation but also has a positive effect on overall performance. This encourages us to explore the power of the CD approach in its simplest form.

The ansatz of CD is not unique. In the Supplementary Material, we compare the performance of 5 different CD ansatz, which are $A = \sum_i \alpha_i J_i\sigma_y^i $(Y-type), $\mathcal{A} = \sum_i \alpha_i J_i \sigma_y^i + \beta \sum_{i<j} J_{ij}\sigma_y^i\sigma_z^j $(Y + YZ${}_\text{u}$)-type), $\mathcal{A} = \sum_i \alpha_i J_i \sigma_y^i + \beta \sum_{i<j}J_{ij}\sigma_y^i\sigma_z^j  +\gamma \sum_{i<j} J_{ij}\sigma_z^i\sigma_y^j$((Y + YZ${}_\text{u}$ + ZY${}_\text{u}$)-type), $\mathcal{A} = \sum_i \alpha_i J_i \sigma_y^i + \sum_{i<j} \beta_{ij}J_{ij}\sigma_y^i\sigma_z^j $((Y + YZ)-type) and $\mathcal{A} = \sum_i \alpha_i J_i \sigma_y^i + \sum_{i<j} \beta_{ij}J_{ij}\sigma_y^i\sigma_z^j +\sum_{i<j} \gamma_{ij}J_{ij}\sigma_z^i\sigma_y^j$((Y + YZ + ZY)-type). It was found that except for the Y-type ansatz, all other ansatz have comparable performances.

\begin{figure*}
\centering
\includegraphics[page = 1, width = 1\textwidth]{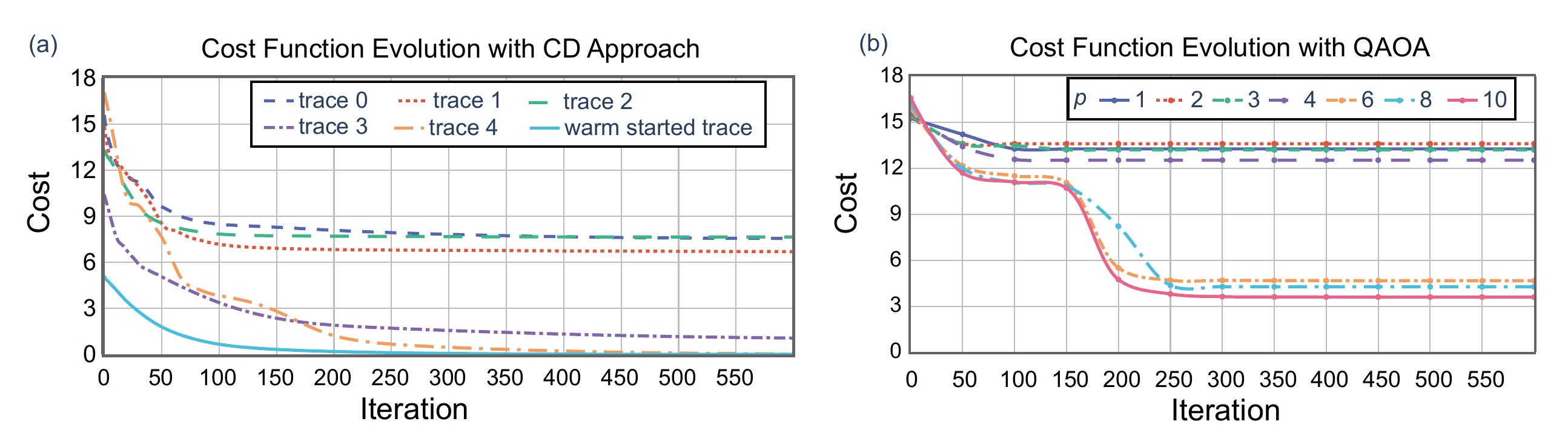}
\caption{Convergent curve for CD approach(a) and QAOA(b) for $H^{9q}$ with cost function being zero at global minimum. The left plot shows a convergent curve with 5 randomly generated initial points and one warm starting point from CD approach with (Y+ ZY${}_\text{u}$)-type ansatz. The right plot shows the convergent curve with initialization inspired by trotterized adiabatic evolution. Different lines correspond to different layer numbers $p$.}
\label{fig2}
\end{figure*}

The accuracy of the algorithm can be further improved by treating the parameters in $A$ as variational and using a classical optimizer to update them, just as in a variational quantum algorithm. Figure~\ref{fig2}(a) shows the learning curve for an optimization process from the CD algorithm with (Y+ ZY${}_\text{u}$)-type CD ansatz. As a comparison, 30 randomly initialized optimization processes are carried out with 5 plotted in the graph. For the random initialization, each $\alpha_j$ is sampled by a Gaussian distribution $\mathcal{N}(1, \sigma_j)$ where $\sigma_j = (2\pi)/(2dt\dot{\lambda}J_i)$ to make sure we sample all regions in the parameter space. The situation of $\beta$ is a bit complicated as it couples to multiple terms, here we set the sampling distribution to be $\mathcal{N}(1,2)$. As is clear from the plot, the warm-starting trace starts with a lower cost and converges faster. Among 30 randomly initialized process, 4 of them (see trace 4 as an example) converges to the global minimum, while all others are stuck at the local minimum. More discussions about the 30 randomly intialized optimization processes have been made in the Supplementary Material. We show that the energy landscape in the parameter space is more complicated than that of the configuration space. Some local minimum corresponds to mixed configuration states.

We will also implement a QAOA solution to the $H^{9q}$ problem. With this approach, the final state is prepared by alternating layers  of mixing, and the problem Hamiltonian layer is expressed as 
\begin{equation}
|\beta, \gamma\rangle = \prod_{i = 1}^p(\exp(-i\beta_i H_i)\exp(-i \gamma_i H_f)|+\rangle^{\otimes n} ,
\end{equation}
with $H_i$ and $H_f$ defined as in Section II. Since QAOA can be viewed as a trotterized version of adiabatic quantum computing (AQC), we assume here a linear scheduling function that ramps Hamiltonian in AQC from $H_i$ to $H_f$, i.e. $\lambda(t) = t/\tau$.  With this, the initialization is chosen such that $\beta_i = \lambda(t_i)/2$ and $1 - \lambda(t_i)$. Classical optimization is then carried out to further reduce the cost function. The convergent curve is presented in Fig.~\ref{fig2}. Different curves correspond to different layers $p$. The global minimum is achieved when the cost function is zero. What the figure displays is that increasing the layer $p$ may lower the final energy, but all cases got stuck in the local minimum. While different initialization may lead to better performance, according to Ref.~\cite{yan2022factoring}, QAOA achieves a 2\% accuracy for a 10-qubit problem with random initialization and $p = 1$.

Similar procedures can be done for $H^{5q}$ and $H^{12q}$. For CD approach with Y + YZ${}_\text{u}$ ansatz, the accuracy is 36.2\% and 5.0\%, respectively. Warm-starting optimization from these CD solutions yields 100\% accuracy for both cases.  

\section{Mixed {\it p}-Spin Problem}
The mixed $p$-spin problem can be defined by
\begin{align}
H = \sum_{k = 1}^p \frac{1}{N^{(p-1)/2}} J_{i_1\ldots i_k} \sigma_{i_1}^z \ldots \sigma_{i_k}^z ,
\end{align}
where $J_{i_1\ldots i_k}$ is random variables following Gaussian distribution $\mathcal{N}(0,1)$. Unlike the factorization problem in Section III, this problem involves all interactions up to order $p$.

This model belongs to the class of random optimization problems and is closely related to other members like K-SAT, q-coloring, p-XORSAT, MaxCut, and Maximal independent set, among others. In many of these systems, the average case presents an algorithmic hardness that prevents stable algorithms from finding optimal or near-optimal solutions. This obstruction for optimality is believed to result from the overlap gap property (OGP)~\cite{gamarnik2021overlap, montanari2021optimization,farhi2020quantum,gamarnik2014limits}, where near-optimal solutions form clusters in the solution space, with forbidden distance between clusters scales with the system size. The OGP can be used to mathematically rule out large classes of algorithms with a form of input stability as potential contenders, including QAOA. One way to circumvent the problem from discontinuity in the solution space to a stable algorithm is to explore the neighbourhood of solutions with warm-starting. This motivates us to test the applicability of the warm-started variational method introduced in Section II.

\begin{figure}
\centering
\includegraphics[page = 1, width = 0.45\textwidth]{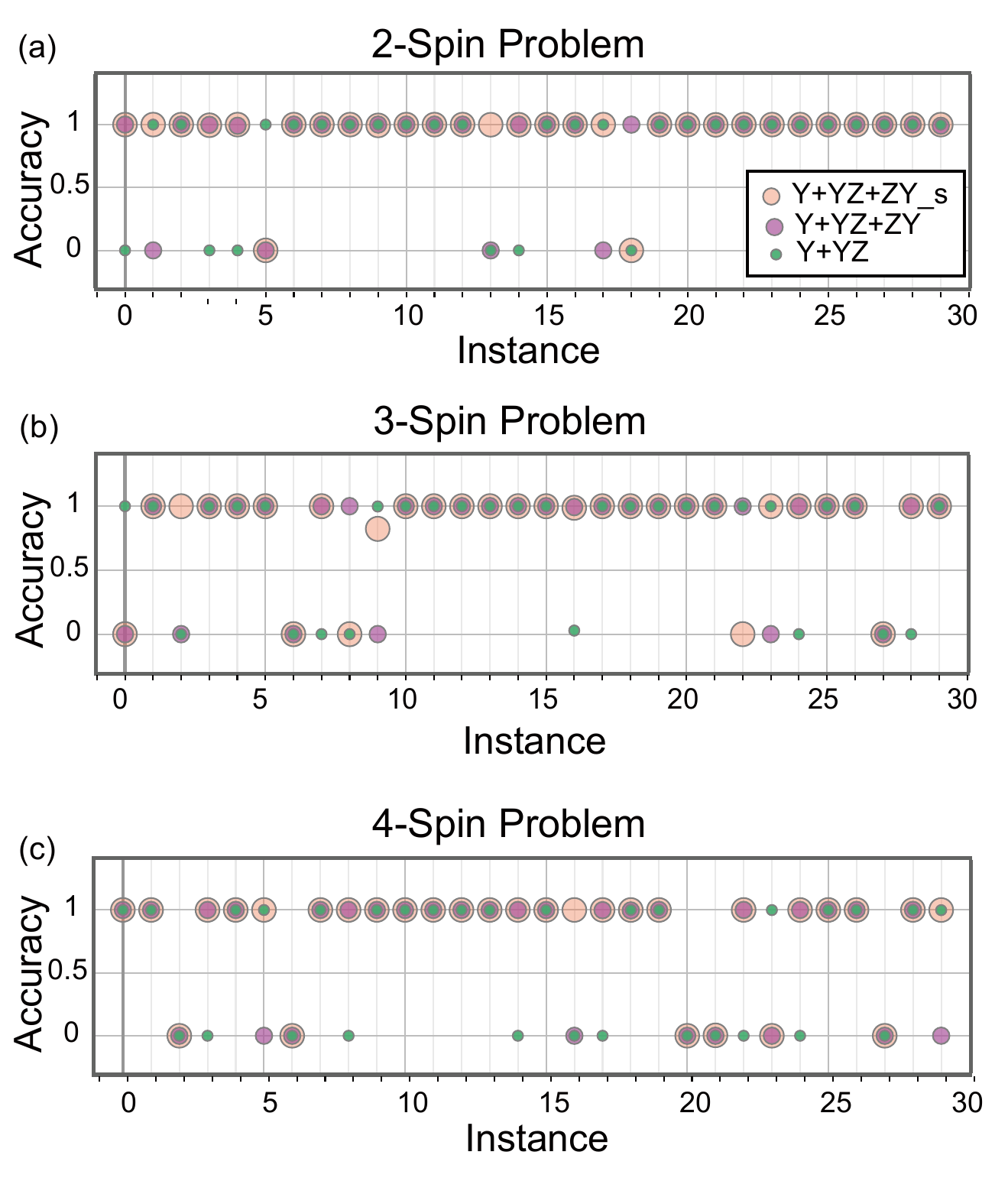}
\caption{Accuracy of warm-starting variational approach for 2-spin(a), 3-spin(b) and 4(c) respectively. Initialization of parameters is determined by minimizing actions as defined in Eq.~\eqref{eq:3}. Dots with different sizes and colors correspond to different CD ansatz as indicated in (a). Definitions of these ansatz are given in the main text. For each $p$-spin problem, 30 instances are randomly generated whose parameters follow Gaussian distribution with zero mean and unit variance. Qubit number is 9.}
\label{fig:p-spin}
\end{figure}
 
We apply the CD approach to 30 randomly generated 2-spin, 3-spin and 4-spin problems with 9 qubits. Three types of CD ansatz are considered, (Y+YZ+ZY)-type, (Y+YZ)-type are defined in the same way as in Section III. We have also introduced another one which denoted by (Y+YZ+ZY${}_s$)-type which is defined as $A = \sum_i \alpha_i J_i\sigma_y^i + \sum_{i<j}\beta_{ij} J_{ij}(\sigma_y^i\sigma_z^j +\sigma_z^i\sigma_y^j)$. This is different from (Y+YZ+ZY${}_u$)-type of ansatz, which is inspired by nested commutator formula. In fact, it could involve 3-local and 4-local CD terms in 3-spin and 4-spin cases, which is beyond the scope of this manuscript. Based on these solutions, optimization of $\alpha$'s and $\beta$'s are made to lower the cost function.

Results are shown in Fig.~\ref{fig:p-spin}, where the x-axis corresponds to different instances and the y-axis describes accuracy after optimizing the parameters. Disk of different colors corresponds to different CD ansatz. As shown in the plots, for the 2-spin problem, there are 21 cases whose ground states are solved by all ansatz, and all cases are solved by at least one ansatz. For the 3-spin problem, there are 18 cases whose ground states are solved by all ansatz and 28 out of 30 solved by at least one ansatz. For the 4-spin problem, half of the instances are solved by all ansatz, and 25 out of 30 are solved by at least one ansatz. From the result, one can tell that the (Y+YZ+ZY${}_s$)-type CD solution produces more good initialization, than the others, but there are also cases when only (Y+YZ+ZY)-type, (Y+YZ)-type CD ansatz works. In general, more parameters imply a closer distance to the true ground state by the CD approach. In the meanwhile, there are more points in the parameter space that correspond to global or local minimum. There is a trade-off between complex and high dimensional parameter space and more smooth lower dimensional space and it may be necessary to try all three ansatz. Since the number of ansatz does not scale with problem size, this does not affect the complexity of the problem. As shown in the plots, 2-local CD terms provide a good enough initialization scheme for the 2-spin problem. However, its ability to capture more complicated higher-order interactions is weakened. Another CD ansatz may be needed to produce a close enough initialization point. Moreover, for the $p$-spin problem, the presence of OGP is only proved for $p\geq 4$. Thus, it is unclear whether the CD approach initialized variational method can combat OGP.

\section{Experimental Results}
Given the simple form of the proposed approach, the method allows the study of the relatively complex system within the constraints of the NISQ era. Implementation involves two practical aspects, exponentiating the CD terms and applying the SWAP strategy to accommodate hardware connectivity. For the former, we use the approach proposed in Ref.~\cite{weidenfeller2022scaling}, decompose into native gates, and merge the YZ component with SWAP gates to further reduce circuit depth. See the Supplementary Material for the construction of the basic building blocks of the algorithm. 

\begin{figure}
\centering
\includegraphics[page = 1, width = 0.45\textwidth]{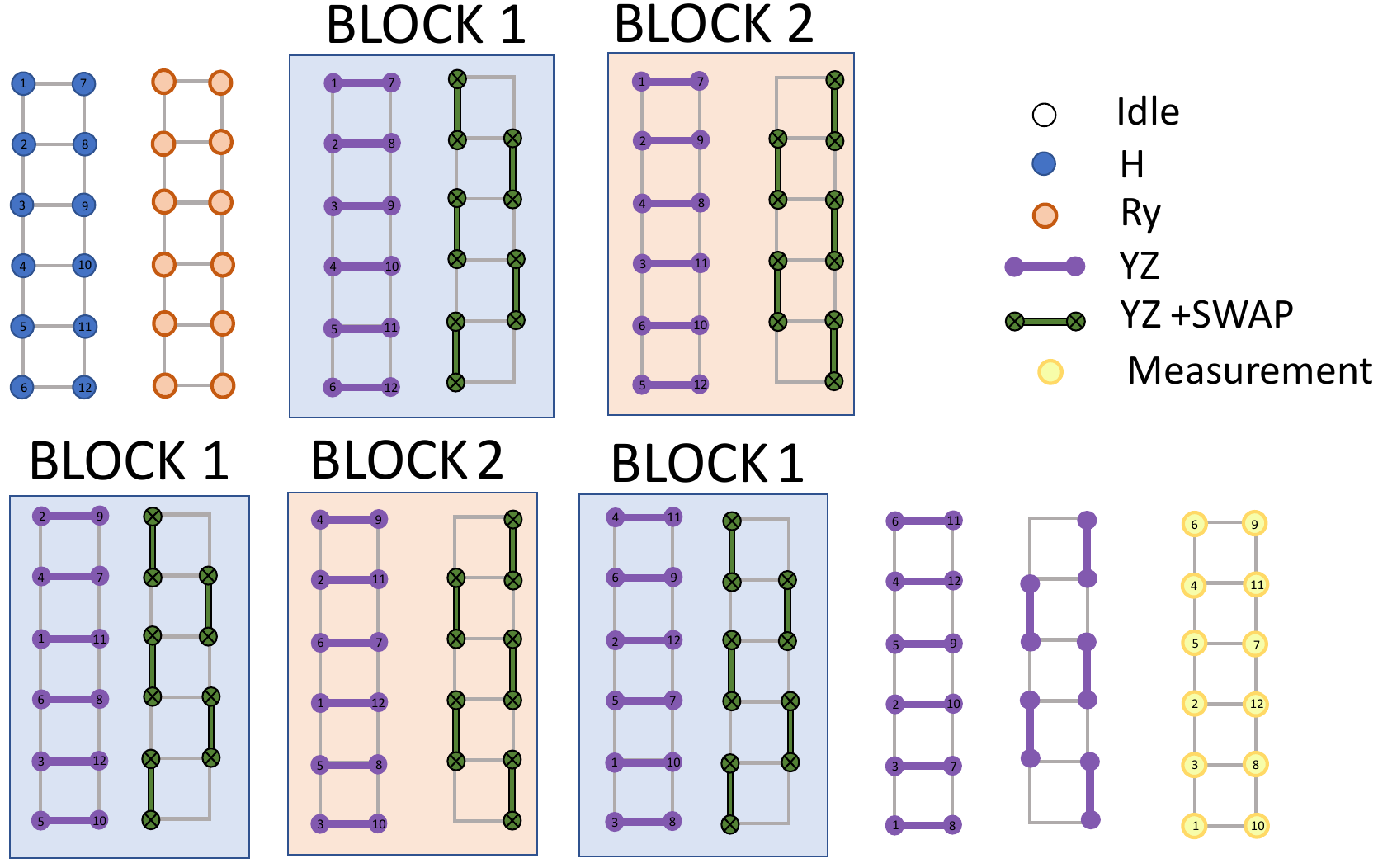}
\caption{Circuit diagram for implementing a 9 qubit swap strategy on a 2 by 6 grid. Circles and dumbbells are representations of gates as listed in the legend. Blocks are categorized by the swap pattern implemented. Numbers on the dots are to track the shuffle among qubits.}
\label{fig:circuit_9}
\end{figure}

For the swap strategy, we adopt the approach in Ref.~\cite{weidenfeller2022scaling}. Based on this method, we studied the influence of topology on circuit depth. It was found that for an even number of qubits, the topology with 2 columns outperforms those with a square grid. Figure~\ref{fig:circuit_9} shows the circuit diagram for implementing a 9 qubit algorithm with (Y+YZ)-type CD ansatz on a 2 by 6 grid. The meanings of the shapes are listed in a legend. As shown in the figure, an H gate followed by a RY gate is applied to each qubit. Then a series of YZ, YZ+SWAP gate is applied for interaction as well as bringing qubits together. In the end, a measurement is made on the whole system. A brief description of the swap strategy as well as a discussion about the shape of the grid to circuit depth are listed in Supplementary Material. We have also included a circuit duration relationship with the qubit number for a CD approach of (Y+YZ)-type in the Supplementary Material, which can be referred to as guidance for approximate problem size accessible by the hardware given a certain coherent time.

\begin{figure}
\centering
\includegraphics[page = 1, width = 0.45\textwidth]{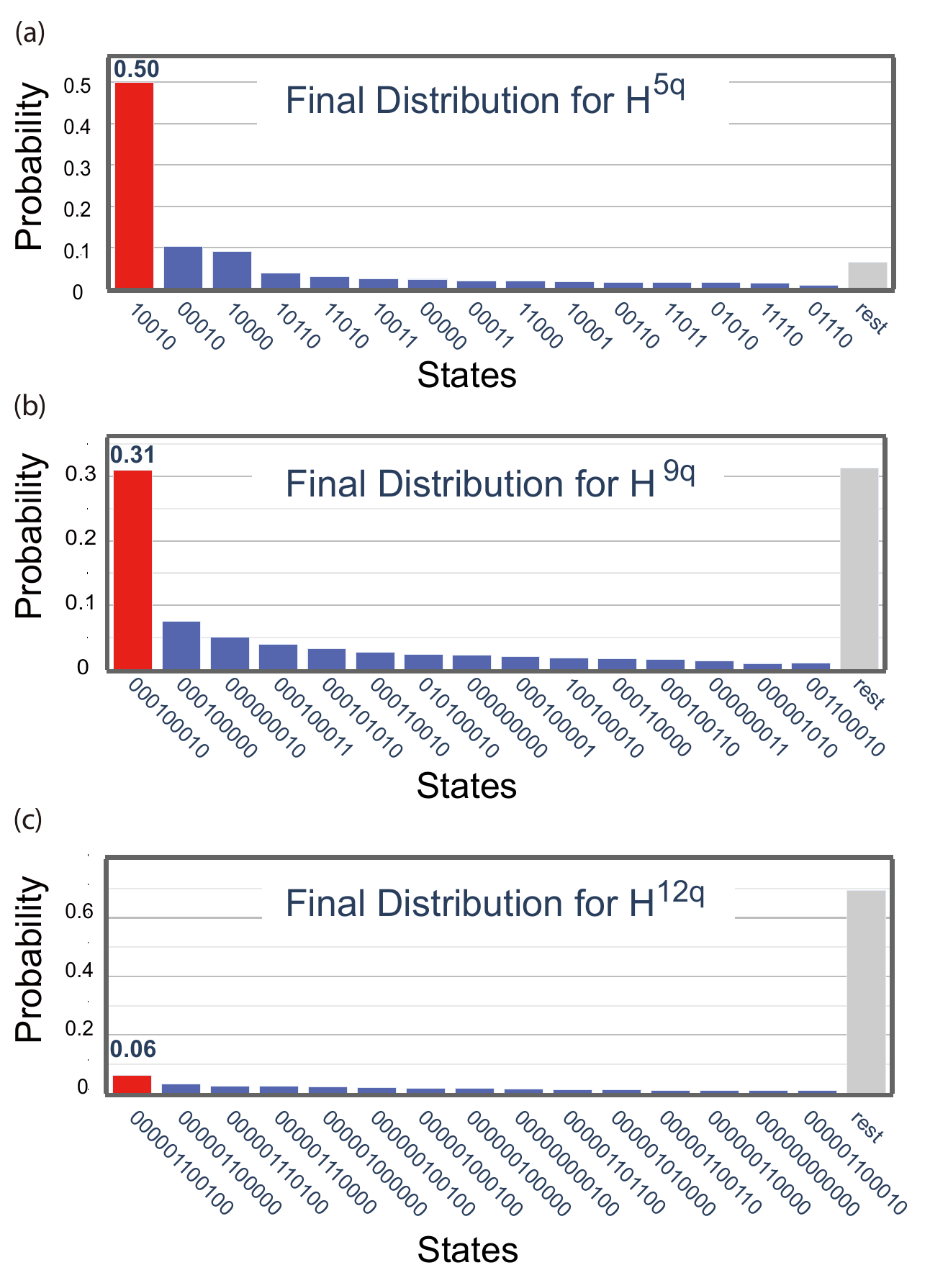}
\caption{Final state distribution after implementing the counterdiabatic approach on the hardware for $H^{5q}$(a), $H^{9q}$(b) and $H^{12q}$(c) respectively. Parameters are first optimized on an ideal simulator and fixed at their optimal values while implementing the algorithms on hardware. The 15 most frequent states in the final distribution are plotted, and others are collected and labeled as the rest. Red bars represent target states, blue ones are for more likely states, and grey ones are for the rest of states.}
\label{fig:6}
\end{figure}

The experimental realization was carried out on a 54-qubit superconducting quantum computer~\cite{huang2020superconducting} from QuantumCtek. See Supplementary Material for processor topology and characterization parameters. Results for factorization problem $H^{9q}$ as well as $H^{5q}$ and $H^{12q}$ are shown in Fig.~\ref{fig:6}. The parameters of the circuit are chosen at optimal values after numerically optimizing the cost function with a warm starting. For ideal simulation, the circuits yield 100\% accuracy for all three cases. When implemented on the hardware, the accuracy drops to 50\%, 31\%, and 6\% for 5-qubit, 9-qubit, and 12-qubit factorization problems, respectively. The decrease in accuracy is a combined effect of gate error and loss of coherence. However, we witness a severe drop for $H^{12q}$ with circuit duration 5.22$\mu s$, which is approaching the limit of coherence on our hardware. That makes it the major source of errors for this experiment. The circuit duration for 5-qubit and 9-qubit algorithms is 2.56$\mu s$ and 4.33$\mu s$, which stay within the coherent time. Thus, limited coherent time becomes the bottleneck for the implementation of complex algorithms on hardware.

\section{Conclusion}
We studied the applicability of a single-layer DCQO paradigm applied to prime factorization and mixed $p$-spin problems. We showed that by an engineered scheduling function as well as fine-tuned $h_x$, a single-layer DCQO circuit can produce important overlap with a global minimum for system size up to 12 qubits for the factorization problem with sparse high-order interaction. Moreover, the proposed quantum algorithm can be further optimized with classical optimizers from the DCQO solution. This approach can solve factorization and 2-spin problems successfully. For 3-spin and 4-spin, a subset of the solution can be solved with variational quantum algorithms from 2-local CD terms. We conjecture that it is related to a phenomenon called ``transient chaos'' in optimization parameters space, where a tiny difference in the space results in a completely diverse optimization path~\cite{achlioptas2006solution}. This phenomenon has been observed in $K$-SAT problem with ($K>3$). Given the trotterized circuit from approximated CD terms inevitably introduces errors, the sensitivity may magnify the effect and lead to wrong convergence. Moreover, we do not claim to solve the OGP problem but provide an improvement by starting in the vicinity of the global minimum. As we have shown, in a highly non-convex parameter space of factorization problem, QAOA is always stuck at a local minimum while the variational approach from CD solutions proves to work for all cases considered with spare high-order interaction. The CD approach provides a highly efficient way to encode high-order interactions into 2-local interactions. This is achieved by solving a variational equation of the action such that the coefficient of the 2-local CD terms keeps information about the whole Hamiltonian. This is valuable for hardware with finite connectivity where swapping strategies for 3-local and 4-local interactions are still missing. This allows us to study systems with high-order interactions without the need to implement complicated Hamiltonians. In the future, it would be interesting to study more types of CD ansatz and explore the efficiency of the algorithm for random optimization problems with different interaction densities such as those defined on c-regular hypergraph or Erd\H{o}s-R\'enyi hypergraph.

\begin{acknowledgments}
 This work is supported by NSFC (12075145), EU FET Open Grant  EPIQUS (899368), Youth Talent Lifting Project (Grant No. 2020-JCJQ-QT-030), National Natural Science Foundation of China (Grant No. 12274464) the Basque Government through Grants No. IT1470-22, the project grants PID2021-126273NB-I00, PID2021-123131NA-I00 funded by MCIN/AEI/10.13039/501100011033, by "ERDF a way of making Europe", and by the European Union NextGenerationEU/PRTR", and the IKUR Strategy of the Basque Government under collaboration of Ikerbasque Foundation and the University of the Basque Country. X.C. acknowledges ayudas para contratos Ram\'on y Cajal–2015-2020 (RYC-2017-22482). F. A.-A. acknowledges to Agencia Nacional de Investigación y Desarrollo: Subvenci\'on a la Instalaci\'on en la Academia SA77210018, Fondecyt Regular 1231174 and Financiamiento Basal para Centros Cient\'ificos y Tecnol\'ogicos de Excelencia AFB 220001.
\end{acknowledgments}

\textbf{Author contributions}
Enrique Solano and Heliang Huang supervised the research. Narendra N. Hegade, Huijie Guan, Xi Chen, and Francisco Albarr\'an-Arriagada developed the theoretical framework. Huijie Guan and Fei Zhou carried out the experiment. Narendra N. Hegade, Huijie Guan, and Fei Zhou analyzed the results. All authors contributed to discussions of the results, development, and conclusions of the manuscript.

\textbf{Conflict of interest}
The authors declare that they have no conflict of interest.

\bibliography{main}

\onecolumngrid

\newpage

\appendix

\begin{center}
\textbf{\large Supplementary Material: Single-Layer Digitized-Counterdiabatic Quantum Optimization for $p$-spin Models}
\end{center}

\section{Introduction} The solution to an optimization problem is affected by several aspects, the hardness of the problem, the efficacy, and the efficiency of the algorithm. Aside from them, practical issues such as compatibility with hardware constraints have also to be accounted for. In this supplementary material, additional discussions will be provided for constructing and characterizing problems~(\ref{sec:factorization}, \ref{sec:p-spin}), evaluation and comparison of approaches~(\ref{sec: cd_approach}), algorithm transpilation~(\ref{sec: transpile}), hardware characterization~(\ref{sec:device}). In section (\ref{sec: transpile}), we present a relation between the system size in qubit number and circuit depth, hoping this could provide guidance for experimental implementation to probe the hardware limit of any device.

\section{Factorization Problem}\label{sec:factorization}

\subsection{Factorization as a constraint equation set}
The use of parametrized quantum circuits for integer factorization was first proposed in Ref.~\cite{anschuetz2019variational}, followed by works in Ref.~\cite{hegade2021digitized}. The so-called variational quantum factoring (VQF) transforms the factorization problem into a set of constraint equations by a series of classical preprocessing steps. To be more specific, each of the two factors $p$ and $q$ are denoted by their binary representations, and the factorization expression $N = p\times q$ produces a set of equations among the binary bits of $p$, $q$ and carry bits $c$. For simplicity, the bit lengths of $p$ and $q$ are known~\textit{a priori}. This allows us to restrict to fewer variables and set the most and least significant digits to be 1. Take factorizing $1261 = 13 \times 97$ as an example, the two factors $p$ and $q$ are represented as $p = (1, p_1, p_2, 1)$ and $q = (1, q_1, q_2, q_3, q_4, q_5, 1)$. Though, more generally, one should remove this pre-assumption and set $n_p$ and $n_q$ according to $n_p = m\left(\lfloor{\sqrt{N}}\rfloor\right) - 1$, $n_q = m\left(\lfloor \frac{N}{3}\rfloor\right) -1 $. Here, $\lfloor a\rfloor$ denotes the greatest odd integer less than or equal to $a$ and $m(b)$ indicates the smallest number of bits required for representing $b$ \cite{peng2008quantum}. With the aforementioned treatment, one obtains $m(N)$ equations (clauses) corresponding to each bit in the factorization equation. To make things easier, we will retain the pre-assumption about the bit-length of $p$ and $q$ in this work.  This is in agreement with those used in Refs.~\cite{anschuetz2019variational, hegade2021digitized}.

\subsection{Simplifying clauses in classical preprocessing}

A crucial step in transforming the factorization problem to an Ising Hamiltonian is to remove trivial solutions due to the boolean nature of the variables. This step is done with classical preprocessing. This work follows the preprocessing steps used in Ref.~\cite{anschuetz2019variational}, with different simplification rules which will be described in this section.

\begin{figure}
\centering
\includegraphics[width = 0.7\textwidth]{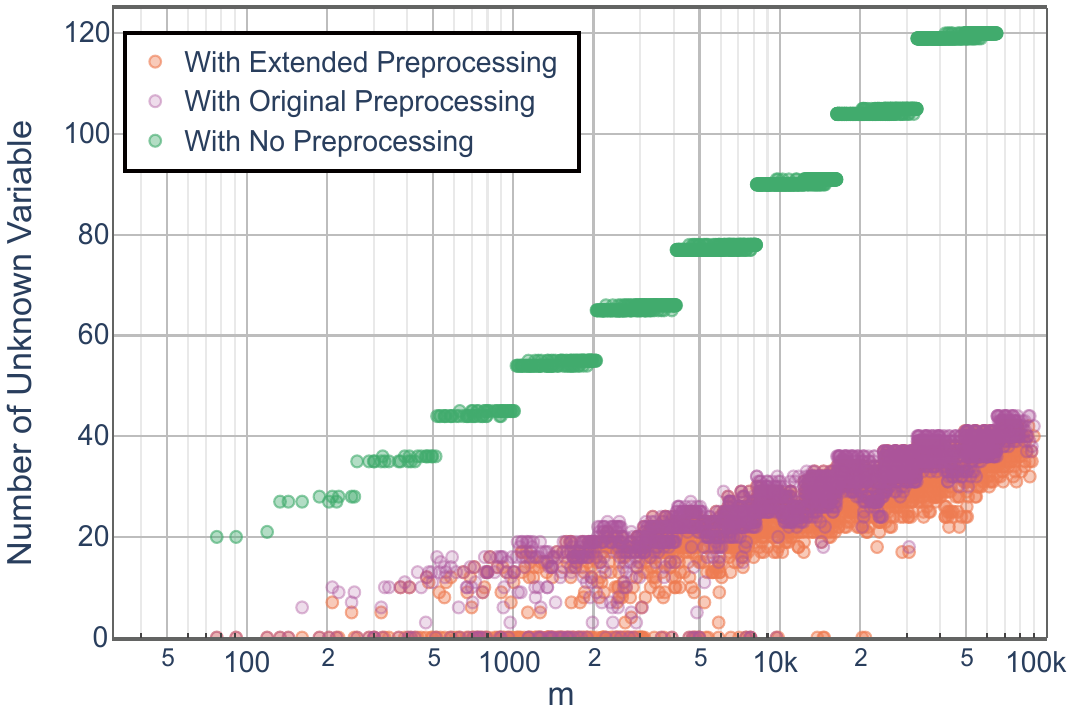}
\caption{Comparison of the number of variables of reduced factorization problem. Red dots related to the case with no preprocessing. Blue dots describe results with original preprocessing as in Ref.~\cite{anschuetz2019variational}. Orange is obtained with new simplification rules with extended preprocessing. More cases become trivial with extended preprocessing, i.e. completely solved with no unknown variables.}
\label{supp_fig1}
\end{figure}

The modification brings the benefit of enhancing reduction while keeping the complexity similar. As shown in Fig.~\ref{supp_fig1}, there are more orange dots (extended preprocessing) on the trivial horizon with no unknown variables than blue dots (original preprocessing). Besides, the cloud of orange dots is slightly lower than that of the original preprocessing results (blue dots). This strips away superficial complexity in some factorization problems, which would otherwise be modeled with more variables but sparser interactions, even decoupled variable sets. This makes them easier to solve. 

The simplification rule we introduced can be categorized into two types, weight rules and parity rules. First write a clause as 
\begin{align}\sum_i a_i f_i(\{p, q, z\}) - \sum_j b_j g_j(\{p, q, z\}) + \text{const}_1 - \text{const}_2 = 0 ,
\end{align}
where $0\leq a_1\leq a_2\leq \ldots\leq a_n$ and $0\leq b_1\leq b_2 \ldots \leq b_m$. Here, $\text{const}_1$($\text{const}_2$) is nonzero if the clause has a positive (negative) constant. If there is no constant term in the clause, then $\text{const}_1 = \text{const}_2 = 0 $. $f_i$ and $g_j$ are products of at most two variables from the sets $\{p, q , c\}$, say $p_1$ or $q_1 c_2$.  Denote the sum of positive coefficients as $A = \sum_i a_i + \text{const}_1$ and the sum of negative coefficients as $B = \sum_j b_i + \text{const}_2$. The weight rules are defined as follows. 


\begin{enumerate}
\item 0 rule

If $a_n > B$ then $f_n(\{p_i, q_i, c_i\}) = 0$. Similarly, if $b_n > A$, then $g_n(\{p, q, z\}) = 0$. E.g. $2p_1 - q_2 * p_3 = 0 \Rightarrow p_1 = 0$
\item 1 rule

If $A - a_n < \text{const}_2$, then all variables appeared in $f_n(\{p, q, z\})$ is 1. Similarly, if $B - b_n <\text{const}_1$, then all variables in $g_n(\{p, q, z\})$ is 1. E.g $3p_1q_2 + q_1 - 2 q_2 - 2 = 0\Rightarrow p_1 = q_2 = 1$
\item compensate rule

If $a_n + a_{n-1} > B$ and $A - (a_n + a_{n-1}) < \text{const}_2$, then $f_n(\{p, q, z\}) + f_{n-1}(\{p, q, z\}) = 1$. Similarly, if $b_n + b_{n-1} > A$ and $B-(b_n + b_{n+1}) < \text{const}_1$, then $g_n(\{p, q, z\}) + g_{n-1}(\{p, q, z\}) = 1$. E.g. $2p_1 + q_2 + q_3  - 2 = 0 \Rightarrow p1 = 1- q_2$

\item identical rule

If $A = B$, $\text{const}_1 = \text{const}_2 = 0$. If set $\{a_i\}$ constains only one element $a$, then $a = b_1 = \ldots b_m$. Similarly, if set $\{b_i\}$ constains only one element $b$, then $ b = a_1= \ldots = a_n$.  E.g. $3p_1 - q_1 - 2 p_2 = 0 \Rightarrow p_1 = p_2 = q_1$
\end{enumerate}

The parity rule extracts out terms with odd coefficients and marks the set by the parity of the constant terms
\begin{enumerate}
\item When the size of the set is 2 and the constant is even, the variables must equal. E.g. $3p_1 + 2p_3 * q_3 - 2q_2 - 3p_2 = 0 \Rightarrow (p_1, p_2)_{\text{even c}} \Rightarrow p_1 = p_2$
\item When the size of the set is 2 and the constant is odd, they must add up to one. E.g. $3p_1 + 2p_3 * q_3 - 2q_2 - 3p_2 -1 = 0 \Rightarrow (p_1, p_2)_{\text{odd c}}\Rightarrow p_1 =  1- p_2$
\item When the size of the set is 1 and the constant is even, the variable must be 0. E.g.
$3p_1 + 2p_3 * q_3 - 2q_2 - 2p_2 - 2 q_1 * p_4 = 0\Rightarrow (p_1)_{\text{even c}} \Rightarrow p_1 = 0$
\item When the size of the set is 1 and the constant is odd, each variable involved must be 1 E.g. $3p_1 + 2p_3 * q_3 - 2q_2 - 2 p_2 - 1= 0 \Rightarrow (p_1)_{\text{odd c}}\Rightarrow p_1 = 1$
\end{enumerate}

The complexity of the algorithm comes from iteratively checking $O(n)$ clauses for the applicability of the rules. Ranking the coefficients for each clause costs $O(n)$ since only the largest two coefficients are needed. The number of iterations is $O(n)$ since the iteration will stop when no new information is generated. Thus the cost is $O(n^3)$ which agrees with that in Ref.~\cite{xu2012quantum}.

With the preprocessing outlined above, the number of variables for factoring 1261, 767, 9983 reduces from 22, 17, 34  to 5, 9, 12. Note that a larger number to factor does not necessarily indicate a more involved problem after preprocessing. Some factorization problems like $9509 = 37 \times 257$ can also be trivial. 

As an example, the clauses for factoring 1261 after simplification read
\begin{align}
p_1 - p_2 + q_5 - 2z_{56} = 0 , \\
p_1q_5 - z_{56} = 0 , \\
p_1 + p_2q_5 + z_{56} - 2z_{78} - 1 = 0 , \\
p_2 + q_5 + z_{78} - 2 = 0 .
\end{align} 

\subsection{Hamiltonian for Factoring 1261, 767, 9983}

As a last step, convert unknown variables to spins with the mapping $x = (1 + \sigma^z)/2$, one obtains the corresponding Ising Hamiltonian. For factoring 1261, 767, 9983, the corresponding Ising Hamiltonian is denoted as $H^{5q}$, $H^{9q}$ and $H^{12q}$, whose expressions are listed in the following. Ground state energy is always 0.

\begin{eqnarray}\label{eq:h5q}
H^{5q} = &&\frac{23}{4}- \frac{5 {\sigma_z^{(0)}}}{4} - \frac{{\sigma_z^{(1)}}}{4} - \frac{5 {\sigma_z^{(2)}}}{4} +\frac{{\sigma_z^{(3)}}}{2} + {\sigma_z^{(4)}}
- \frac{{\sigma_z^{(0)}} {\sigma_z^{(1)}}}{4} + \frac{3 {\sigma_z^{(0)}} {\sigma_z^{(2)}}}{4} - \frac{3 {\sigma_z^{(0)}} {\sigma_z^{(3)}}}{4} - {\sigma_z^{(0)}} {\sigma_z^{(4)}} - \frac{{\sigma_z^{(1)}} {\sigma_z^{(2)}}}{4} \nonumber\\
 &&+ \frac{5 {\sigma_z^{(1)}} {\sigma_z^{(3)}}}{4} - {\sigma_z^{(2)}} {\sigma_z^{(3)}} - {\sigma_z^{(3)}} {\sigma_z^{(4)}}
+\frac{{\sigma_z^{(0)}} {\sigma_z^{(1)}} {\sigma_z^{(2)}}}{4} -\frac{{\sigma_z^{(0)}} {\sigma_z^{(2)}} {\sigma_z^{(3)}}}{4} +\frac{{\sigma_z^{(1)}} {\sigma_z^{(2)}} {\sigma_z^{(3)}}}{4} - \frac{{\sigma_z^{(1)}} {\sigma_z^{(2)}} {\sigma_z^{(4)}}}{2} ,
\end{eqnarray}

\begin{eqnarray}\label{eq:h9q}
H^{9q} = && \frac{121}{8} + \frac{5 {\sigma_z^{(0)}}}{8} + \frac{{\sigma_z^{(1)}}}{8} - \frac{11 {\sigma_z^{(2)}}}{8} - \frac{15 {\sigma_z^{(3)}}}{8} - \frac{{\sigma_z^{(4)}}}{4} - {\sigma_z^{(5)}} + \frac{11 {\sigma_z^{(6)}}}{4}+ 5 {\sigma_z^{(7)}} - \frac{{\sigma_z^{(8)}}}{2}
+\frac{{\sigma_z^{(0)}} {\sigma_z^{(1)}}}{8} - \frac{3 {\sigma_z^{(0)}} {\sigma_z^{(2)}}}{8} \nonumber\\
&&+ \frac{5 {\sigma_z^{(0)}} {\sigma_z^{(3)}}}{8} + \frac{{\sigma_z^{(0)}} {\sigma_z^{(4)}}}{4} 
 - \frac{3 {\sigma_z^{(0)}} {\sigma_z^{(5)}}}{4} + {\sigma_z^{(0)}} {\sigma_z^{(7)}} + {\sigma_z^{(0)}} {\sigma_z^{(8)}} + \frac{5 {\sigma_z^{(1)}} {\sigma_z^{(2)}}}{8} - \frac{3 {\sigma_z^{(1)}} {\sigma_z^{(3)}}}{8} + \frac{{\sigma_z^{(1)}} {\sigma_z^{(5)}}}{4} 
- \frac{3 {\sigma_z^{(1)}} {\sigma_z^{(6)}}}{4}\nonumber\\ 
&&- \frac{3 {\sigma_z^{(1)}} {\sigma_z^{(7)}}}{2} + \frac{5 {\sigma_z^{(2)}} {\sigma_z^{(3)}}}{8} - \frac{3 {\sigma_z^{(2)}} {\sigma_z^{(4)}}}{4} - \frac{{\sigma_z^{(2)}} {\sigma_z^{(5)}}}{4} - {\sigma_z^{(2)}} {\sigma_z^{(7)}} - {\sigma_z^{(2)}} {\sigma_z^{(8)}} + \frac{{\sigma_z^{(3)}} {\sigma_z^{(4)}}}{2} - \frac{3 {\sigma_z^{(3)}} {\sigma_z^{(5)}}}{4} - \frac{{\sigma_z^{(3)}} {\sigma_z^{(6)}}}{4}\nonumber\\
&& - \frac{{\sigma_z^{(3)}} {\sigma_z^{(7)}}}{2} - {\sigma_z^{(4)}} {\sigma_z^{(5)}} - {\sigma_z^{(5)}} {\sigma_z^{(6)}} - 2 {\sigma_z^{(5)}} {\sigma_z^{(7)}} + 4 {\sigma_z^{(6)}} {\sigma_z^{(7)}} - {\sigma_z^{(6)}} {\sigma_z^{(8)}} + \frac{{\sigma_z^{(7)}} {\sigma_z^{(8)}}}{2} - \frac{3 {\sigma_z^{(0)}} {\sigma_z^{(1)}} {\sigma_z^{(2)}}}{8} + \frac{{\sigma_z^{(0)}} {\sigma_z^{(1)}} {\sigma_z^{(3)}}}{8}\nonumber\\
&& + {\sigma_z^{(0)}} {\sigma_z^{(1)}} {\sigma_z^{(4)}} - \frac{3 {\sigma_z^{(0)}} {\sigma_z^{(2)}} {\sigma_z^{(3)}}}{8} - \frac{{\sigma_z^{(0)}} {\sigma_z^{(2)}} {\sigma_z^{(4)}}}{4} + \frac{{\sigma_z^{(0)}} {\sigma_z^{(2)}} {\sigma_z^{(5)}}}{2}
 - \frac{{\sigma_z^{(0)}} {\sigma_z^{(3)}} {\sigma_z^{(5)}}}{4} + \frac{{\sigma_z^{(0)}} {\sigma_z^{(3)}} {\sigma_z^{(6)}}}{2} + {\sigma_z^{(0)}} {\sigma_z^{(3)}} {\sigma_z^{(7)}} \nonumber\\
 &&- \frac{3 {\sigma_z^{(1)}} {\sigma_z^{(2)}} {\sigma_z^{(3)}}}{8} - \frac{{\sigma_z^{(1)}} {\sigma_z^{(2)}} {\sigma_z^{(5)}}}{4} 
+ \frac{{\sigma_z^{(1)}} {\sigma_z^{(2)}} {\sigma_z^{(6)}}}{2} + {\sigma_z^{(1)}} {\sigma_z^{(2)}} {\sigma_z^{(7)}} - \frac{{\sigma_z^{(1)}} {\sigma_z^{(3)}} {\sigma_z^{(6)}}}{4} + \frac{{\sigma_z^{(1)}} {\sigma_z^{(3)}} {\sigma_z^{(8)}}}{2} \nonumber\\
&&+ \frac{{\sigma_z^{(0)}} {\sigma_z^{(1)}} {\sigma_z^{(2)}} {\sigma_z^{(3)}}}{8} ,
\end{eqnarray}

\begin{eqnarray}\label{eq:h12q}
H^{12q} = && \frac{117}{8}- \frac{7 {\sigma_z^{(1)}}}{8} - \frac{{\sigma_z^{(2)}}}{4} - \frac{3 {\sigma_z^{(3)}}}{2} + \frac{{\sigma_z^{(4)}}}{8} + \frac{{\sigma_z^{(5)}}}{2} + \frac{{\sigma_z^{(6)}}}{4} + \frac{3 {\sigma_z^{(7)}}}{2} + \frac{{\sigma_z^{(8)}}}{4}- \frac{3 {\sigma_z^{(10)}}}{2} + \frac{3 {\sigma_z^{(11)}}}{2}+\frac{{\sigma_z^{(0)}} {\sigma_z^{(2)}}}{8} - \frac{{\sigma_z^{(0)}} {\sigma_z^{(3)}}}{8} \nonumber\\
&&- \frac{{\sigma_z^{(0)}} {\sigma_z^{(4)}}}{2} + \frac{{\sigma_z^{(0)}} {\sigma_z^{(6)}}}{4} - \frac{{\sigma_z^{(0)}} {\sigma_z^{(7)}}}{2} - \frac{{\sigma_z^{(0)}} {\sigma_z^{(8)}}}{2} + \frac{3 {\sigma_z^{(0)}} {\sigma_z^{(9)}}}{2} - {\sigma_z^{(0)}} {\sigma_z^{(10)}} + \frac{{\sigma_z^{(1)}} {\sigma_z^{(2)}}}{2} + \frac{{\sigma_z^{(1)}} {\sigma_z^{(3)}}}{4} + \frac{{\sigma_z^{(1)}} {\sigma_z^{(4)}}}{8} + \frac{{\sigma_z^{(1)}} {\sigma_z^{(7)}}}{4} \nonumber\\
&&- \frac{{\sigma_z^{(1)}} {\sigma_z^{(8)}}}{2} - \frac{{\sigma_z^{(1)}} {\sigma_z^{(9)}}}{2}+ \frac{3 {\sigma_z^{(1)}} {\sigma_z^{(10)}}}{2}- {\sigma_z^{(1)}} {\sigma_z^{(11)}} + \frac{5 {\sigma_z^{(2)}} {\sigma_z^{(3)}}}{8} + \frac{5 {\sigma_z^{(2)}} {\sigma_z^{(4)}}}{4} - \frac{3 {\sigma_z^{(2)}} {\sigma_z^{(5)}}}{4} - \frac{{\sigma_z^{(2)}} {\sigma_z^{(6)}}}{4} - \frac{{\sigma_z^{(2)}} {\sigma_z^{(7)}}}{4}\nonumber\\
&& - \frac{{\sigma_z^{(2)}} {\sigma_z^{(8)}}}{4} - \frac{{\sigma_z^{(2)}} {\sigma_z^{(9)}}}{2}+ \frac{{\sigma_z^{(2)}} {\sigma_z^{(11)}}}{2} + \frac{{\sigma_z^{(3)}} {\sigma_z^{(4)}}}{2} - \frac{3 {\sigma_z^{(3)}} {\sigma_z^{(5)}}}{4} - \frac{{\sigma_z^{(3)}} {\sigma_z^{(6)}}}{4} - \frac{{\sigma_z^{(3)}} {\sigma_z^{(7)}}}{4} - \frac{{\sigma_z^{(3)}} {\sigma_z^{(8)}}}{2} + \frac{{\sigma_z^{(3)}} {\sigma_z^{(10)}}}{2} \nonumber\\
&&- {\sigma_z^{(3)}} {\sigma_z^{(11)}} + \frac{{\sigma_z^{(4)}} {\sigma_z^{(5)}}}{2} - \frac{3 {\sigma_z^{(4)}} {\sigma_z^{(6)}}}{4} - \frac{{\sigma_z^{(4)}} {\sigma_z^{(7)}}}{4} - \frac{{\sigma_z^{(4)}} {\sigma_z^{(8)}}}{4}  - \frac{{\sigma_z^{(4)}} {\sigma_z^{(9)}}}{2} + \frac{{\sigma_z^{(4)}} {\sigma_z^{(11)}}}{2} - {\sigma_z^{(5)}} {\sigma_z^{(6)}} - {\sigma_z^{(6)}} {\sigma_z^{(7)}} - {\sigma_z^{(7)}} {\sigma_z^{(8)}} \nonumber\\
&& - {\sigma_z^{(8)}} {\sigma_z^{(9)}} - {\sigma_z^{(9)}} {\sigma_z^{(10)}}  - {\sigma_z^{(10)}} {\sigma_z^{(11)}}- \frac{5 {\sigma_z^{(0)}} {\sigma_z^{(1)}} {\sigma_z^{(2)}}}{8} - \frac{3 {\sigma_z^{(0)}} {\sigma_z^{(1)}} {\sigma_z^{(3)}}}{8} + {\sigma_z^{(0)}} {\sigma_z^{(1)}} {\sigma_z^{(5)}} - \frac{3 {\sigma_z^{(0)}} {\sigma_z^{(2)}} {\sigma_z^{(4)}}}{8} - \frac{{\sigma_z^{(0)}} {\sigma_z^{(2)}} {\sigma_z^{(5)}}}{4} \nonumber\\ 
&&+ \frac{{\sigma_z^{(0)}} {\sigma_z^{(2)}} {\sigma_z^{(6)}}}{2} + \frac{3 {\sigma_z^{(0)}} {\sigma_z^{(3)}} {\sigma_z^{(4)}}}{8} + \frac{{\sigma_z^{(0)}} {\sigma_z^{(3)}} {\sigma_z^{(5)}}}{4} - \frac{{\sigma_z^{(0)}} {\sigma_z^{(3)}} {\sigma_z^{(6)}}}{2}+ \frac{{\sigma_z^{(0)}} {\sigma_z^{(4)}} {\sigma_z^{(6)}}}{4} - \frac{{\sigma_z^{(0)}} {\sigma_z^{(4)}} {\sigma_z^{(7)}}}{2} + \frac{{\sigma_z^{(1)}} {\sigma_z^{(2)}} {\sigma_z^{(3)}}}{8}\nonumber\\
&& - \frac{{\sigma_z^{(1)}} {\sigma_z^{(2)}} {\sigma_z^{(6)}}}{4} + \frac{{\sigma_z^{(1)}} {\sigma_z^{(2)}} {\sigma_z^{(7)}}}{2} + \frac{{\sigma_z^{(1)}} {\sigma_z^{(3)}} {\sigma_z^{(4)}}}{4} + \frac{{\sigma_z^{(1)}} {\sigma_z^{(3)}} {\sigma_z^{(6)}}}{4} - \frac{{\sigma_z^{(1)}} {\sigma_z^{(3)}} {\sigma_z^{(7)}}}{2} + \frac{{\sigma_z^{(1)}} {\sigma_z^{(4)}} {\sigma_z^{(7)}}}{4} - \frac{{\sigma_z^{(1)}} {\sigma_z^{(4)}} {\sigma_z^{(8)}}}{2}\nonumber\\
&& + \frac{{\sigma_z^{(2)}} {\sigma_z^{(3)}} {\sigma_z^{(4)}}}{8} +\frac{{\sigma_z^{(2)}} {\sigma_z^{(3)}} {\sigma_z^{(7)}}}{4} - \frac{{\sigma_z^{(2)}} {\sigma_z^{(3)}} {\sigma_z^{(8)}}}{2} + \frac{{\sigma_z^{(2)}} {\sigma_z^{(4)}} {\sigma_z^{(8)}}}{4} - \frac{{\sigma_z^{(2)}} {\sigma_z^{(4)}} {\sigma_z^{(9)}}}{2}- \frac{{\sigma_z^{(0)}} {\sigma_z^{(1)}} {\sigma_z^{(2)}} {\sigma_z^{(4)}}}{8} \nonumber\\
&&+ \frac{{\sigma_z^{(0)}} {\sigma_z^{(1)}} {\sigma_z^{(3)}} {\sigma_z^{(4)}}}{8} + \frac{{\sigma_z^{(1)}} {\sigma_z^{(2)}} {\sigma_z^{(3)}} {\sigma_z^{(4)}}}{8} .
\end{eqnarray}

From these expressions, it can be seen that the Ising problem for factorization problem consists of interactions up to 4-local terms, which holds for any factorization problem. The interaction density for 3-local and 4-local varies from case to case. $H^{5q}$ has 4 out of all 10 possible 3-local interactions and no 4-local interactions present. $H^{9q}$ has 15 out of all 84 possible 3-local interactions 1 out of 126 4-local interactions and 3 out of all 495 possible 4-local interactions. $H^{12q}$ has 12 out of all 220 3-local interactions. In general, Ising spins with low $\gamma$-parameter(number of interactions per variable) are easier. As discussed in Ref.~\cite{mzard2002alternative}, with increasing $\gamma$-parameters, solutions form clusters in the configuration space and the Overlap Gap property will emerge.

\section{Mixed {\it p}-Spin Problem}\label{sec:p-spin}

\subsection{Definition of mixed \texorpdfstring{$p$}-spin}
A $p$-Spin problem describes the optimal or near-optimal solutions to the following Ising Hamiltonian
\begin{align}
H_p = \frac{1}{n^{(p+1)/2}}\langle Y, \sigma_z^{\otimes p}\rangle ,
\end{align}
where $Y$ is a random coefficient tensor of order $p$ with i.i.d $\mathcal{N}(0,1)$ entries. $\sigma_i^z$ are Pauli-$z$ matrices acting on qubit $i$. The mixed $p$-Spin problem is a linear combination,
\begin{align}
H = \sum_{p\geq 1}\beta_p H_p ,
\end{align}
with coefficients $\beta_p$ decreasing fast enough, for example $\sum_{p\geq 1}2^p \beta_p^2<\infty$ to ensure finiteness of energy. We consider the case where $\beta_p = 0$ for $p>4$, i.e.
\begin{align}\label{eq:1}
H = J_0 + \sum_{i} J_{i} \sigma_i^z + \sum_{i<j} J_{ij} \sigma_i^z\sigma_j^z \sum_{i<j<k} J_{ijk}\sigma_i^z\sigma_j^z \sigma_k^z + \sum_{i<j<k<l}J_{ijkl}\sigma_i^z\sigma_j^z\sigma_k^z\sigma_k^l .
\end{align}
Here, $J_{i_1\ldots i_p}$ are random variables following $\mathcal{N}(0,1/N^{(p-1)/2})$. The Hamiltonian is of broad interest for its close relationship to multiple problems which has been the subject of extensive investigation. This is the case of fully connected pure p-spin model ~\cite{gross1984simplest, castellani2005spin}, $p$-XORSAT model~\cite{mezard2003two}\footnote{In a $p$-XORSAT problem, there are $M=\alpha N$ interactions with couplings of order 1, defined on an End\H{o}s-R\'enyi random graph. Control parameter $\alpha$ describes the constraint density. In a fully connected $p$-spin problem, all possible tuples of interactions are present, with individual strength diminishing as $1/\sqrt{N^{(p-1)/2}}$}, and $K$-SAT~\cite{mezard2009information}, among others, when the problem parameter $p/K$ is set to be 4 or less. The Hamiltonian can also be reduced to more familiar ones with pair-wise interactions when the 3-body and 4-body interactions vanish, thus naturally supporting solutions to MaxCut~\cite{farhi2014quantum}, Sherrington-Kirkpatrick (SK) model~\cite{farhi2022quantum}, Maximal independent set~\cite{bomze1999maximum}, q-coloring~\cite{mezard2009information}, among others. 
 
For some problems like the SK model(pure 2-spin problem), the task of solving the minimization problem can be solved successfully~\cite{montanari2021optimization}. For many others, there exhibits an apparent algorithmic hardness. As explained in~Ref.~\cite{gamarnik2021overlap}, using the nonconstructive analysis method, one shows that the optimal value of some optimization problem is some value $c^*$. However, the best known polynomial time algorithm can only achieve value $c_{\text{ALG}} = \rho c^*$, which is strictly less than $c^*$. The presence of a multiplicative gap $\rho<1$ happens in many system mentioned above like $p$-spin~\cite{el2021optimization}, $K$-SAT~\cite{achlioptas2006solution,mezard2005clustering}, maximum independent set~\cite{gamarnik2014limits,rahman2017local}, q-coloring~\cite{achlioptas2006solution}, with some others like Number Partitioning problem~\cite{gamarnik2021overlap}, etc. This obstruction for optimality is believed to result from the Overlap Gap Property (OGP) which manifests itself as a barrier for local algorithms~\cite{gamarnik2021overlap, montanari2021optimization,farhi2020quantum,gamarnik2014limits}.

\subsection{Overlap Gap Property}
 
An Overlap Gap Property (OGP) describes a nontrivial geometry of the near-optimal solutions of the underlying optimization problem. It is closely related to a phenomenon called clustering of a nearly optimal solution, though there are nuances between the two, see Refs.~\cite{gamarnik2021overlap,huang2022tight}. It is observed that for a particular instance of large random graphs problem, any two solutions $\sigma$, $\tau$ that is $\mu$-optimal, i.e. energy is $\mu$ above the optimal value, have overlap which is either very small or very large, i.e. 
 $\langle \sigma|\tau\rangle > \nu_1$ or  $\langle \sigma|\tau\rangle < \nu_2$ with $0<\nu_1 < \nu_2<1$. That is to say, all near-optimal solutions form clusters in the solution space. Moreover, it is observed that the forbidden region between the two solutions has a Hamming distance that scales with the size of the problem.

For problems with OGP, solving for optimal solutions exhibits an apparent algorithmic hardness. As explained in Ref.~\cite{gamarnik2021overlap}, using the nonconstructive analysis method, one shows that the optimal value of some optimization problem is some value $c^*$. However, the best known polynomial time algorithm can only achieve value $c_{\text{ALG}}$, which is strictly less than $c^*$. This obstruction for optimality is believed to result from the Overlap Gap Property (OGP) which manifests itself as a barrier for local algorithms~\cite{gamarnik2021overlap, montanari2021optimization,farhi2020quantum,gamarnik2014limits}

The OGP has been rigorously established for mixed $p$-spin model~\cite{huang2022tight}, $p$-spin model\cite{chen2019suboptimality}, $K$-SAT with $K>8$\cite{achlioptas2006solution,mezard2005clustering}, maximum independent set model~\cite{gamarnik2014limits,coja2015independent}, among others. It is also worth remarking that the OGP is conjectured to not exist for the low energy configurations of the SK model~\cite{montanari2021optimization}.

The OPG can be used to mathematically rule out large classes of algorithms as potential contenders, specifically algorithms exhibiting a form of input stability. This includes both classical algorithms like local algoirthms~\cite{gamarnik2014limits,gamarnik2014limits,chen2019suboptimality}, Markov Chain Monte Carlo (and related) methods~\cite{coja2017walksat,ben2018spectral,gamarnik2021overlapsubmx} and Approximate Message Passing type algorithms~\cite{gamarnik2021overlapMPA} as well as quantum algorithms like quantum approximate optimization algorithms(QAOA) when the circuit depth $p$ is less than a problem-dependent constant times $\log(n)$~\cite{farhi2020quantum}.

\section{Counterdiabatic Approach}\label{sec: cd_approach}
\subsection{Principle of CD Approach} 
Long before the consideration of counterdiabatic evolution, adiabatic processes have been treated as a universal way to steer systems along their eigenstates. As formulated in Ref.~\cite{messiah1964quantum}, evolving under a time-dependent Hamiltonian $\mathcal{H}$ without level crossing, a system will follow the instantaneous eigenstate for long enough total evolution time $\tau$. A more precise description of the adiabatic condition states that
\begin{align}
\tau > \frac{1}{\epsilon}\frac{b(t)}{\Delta(t)^2} , \hspace{20pt}
b(t) = \left|\langle 1|\frac{d\mathcal{H}(t)}{dt}|0\rangle \right| , \hspace{20pt}
\Delta(t) = E_1(t)- E_0(t) .
\end{align}
Here, $b(t)$ describes the transition probability due to the changing Hamiltonian and $\Delta(t)$ is the instantaneous energy gap, $\epsilon^2$ is the probability that the final state fails to follow the eigenstate. 

The adiabatic theorem is the backbone of quantum annealing and adiabatic quantum algorithms. The ground state of a target Hamiltonian can be prepared by evolving the system under a time-dependent Hamiltonian
\begin{align}\label{eq:h_ad2}
\mathcal{H}(\lambda(t)) = (1- \lambda(t))H_i + \lambda(t) H_f
\end{align}
from the ground state of $H_i$. In this sense, $H_i$ is an easily prepared Hamiltonian that does not commute with $H_f$. Usually, a transverse field is set and the initial state is prepared as $|-\rangle^{\otimes n}$. 

To be implemented on a universal quantum computer, trotterization is used to convert the time evolution operator into products of exponential over small time steps. The slowness required by the adiabatic condition then translates into a deep circuit, making the algorithm vulnerable to decoherence, a serious constraint in the NISQ era. Understanding the factors that induce excitation would provide solutions to shortcut the adiabatic evolution at finite time. 

As first proposed in Ref.~\cite{demirplak2003adiabatic}, excitation of states is caused by an emerging gauge potential term in the space spanned by a time-dependent eigenbasis. To see this, denote $|\psi\rangle$ the states in the lab frame with a nonzero off-diagonal matrix with $\mathcal{H}(\lambda)$ and $|\tilde{\psi}\rangle = U(\lambda)|\psi\rangle$ which is an eigenstate of $\mathcal{H}(\lambda)$, $U(\lambda)$ is the transformation from $|\psi\rangle$ to $|\tilde{\psi}\rangle$. Then the Schr\"{o}dinger equations reads
\begin{align}
i\hbar \partial_t|\tilde{\psi}\rangle = \left(\tilde{\mathcal{H}}(\lambda) - \dot{\lambda}\tilde{A}_\lambda\right)|\tilde{\psi}\rangle\hspace{20pt}\text{with}\hspace{20pt}
\tilde{\mathcal{H}}(\lambda) = U^\dagger \mathcal{H}(\lambda) U , \hspace{20pt}
\tilde{A}_\lambda = i U\partial_\lambda U^\dagger ,
\end{align}
where $\tilde{\mathcal{H}}(\lambda)$($\tilde{A}_\lambda$) is $\mathcal{H}(\lambda)$($A_\lambda$) in the eigenbasis frame. $\tilde{A}$ is a gauge potential with off-diagonal elements in the eigenbasis $|\tilde{n}(\lambda)\rangle$, thus causing excitations. The matrix element of $A$ can be evaluated as
\begin{align}
A_\lambda = (i\partial_\lambda U^\dagger)U = -i U^\dagger \partial_\lambda U , \end{align}
\begin{align}\label{eq:gauge_A_16}
\langle n| A|n \rangle = i\hbar\langle \tilde{n}(\lambda)|\partial_\lambda|\tilde{n}(\lambda)\rangle .
\end{align}
Moreover, as $\tilde{\mathcal{H}}$ is diagonal in $|\tilde{n}(\lambda)\rangle$, we have $\langle \tilde{m}(\lambda) |\tilde{\mathcal{H}}|\tilde{n}(\lambda)\rangle = 0$ for $m\neq n$. Thus, differentiating both sides by $\lambda$ yields
\begin{align}\label{eq: gauge_A}
0 = \langle \partial_\lambda\tilde{m}(\lambda) |\tilde{\mathcal{H}}|\tilde{n}(\lambda)\rangle + \langle \tilde{m}(\lambda) |\partial_\lambda \tilde{\mathcal{H}}|\tilde{n}(\lambda)\rangle + \langle \tilde{m}(\lambda) |\tilde{\mathcal{H}}|\partial_\lambda \tilde{n}(\lambda)\rangle .
\end{align}\label{eq:gauge_A_18}

\noindent Since $\mathcal{H}|\tilde{n}(\lambda)\rangle = E_n|\tilde{n}\rangle$, $|\partial_\lambda\tilde{n}(\lambda)\rangle = \partial_\lambda U(\lambda)|n\rangle = iUA|n\rangle$, Eq.~\eqref{eq: gauge_A} leads to 
\begin{align}\label{eq:gauge-compare}
\langle m|A_\lambda|n\rangle = i\hbar\frac{\langle m |\partial_\lambda H |n\rangle}{E_n- E_m} . \hspace{20pt} 
\end{align}
Combining Eqs.~\eqref{eq:gauge_A_16} and \eqref{eq:gauge_A_18}, one can get the matrix form
\begin{align}\label{eq: gauge_A_19}
\partial_\lambda \tilde{\mathcal{H}} = -\frac{i}{\hbar} [\tilde{A}_\lambda, \tilde{\mathcal{H}}] - \tilde{F}_{\text{ad}} \hspace{20pt}
\tilde{F}_{\text{ad}} = \sum_n \partial_\lambda E_n(\lambda)|\tilde{n}(\lambda)\rangle\langle \tilde{n}(\lambda)| .
\end{align}
Here, $\tilde{F}_{\text{ad}}$ is the generalized force, which is diagonal in $|\tilde{n}(\lambda)\rangle$, thus commuting with $\tilde{\mathcal{H}} = \sum_n E_n(\lambda)|\tilde{n}(\lambda)\rangle \langle\tilde{n}(\lambda)|$. This gives rise to a useful property that 
\begin{align}
[\mathcal{H}, i\hbar\partial_\lambda \mathcal{H} - [A_\lambda, \mathcal{H}]] = 0 ,
\end{align}
which holds for both eigen-basis frame $|\tilde{n}(\lambda)\rangle$ and lab frame $|n\rangle$.

Realizing the presence of gauge potential $A_\lambda$ which drives excitations, one can add counterdiabatic (CD) terms that exactly cancel its contribution, i.e. 
\begin{align}
H_{\text{CD}} = \mathcal{H} + \dot{\lambda} A_\lambda .
\end{align}
This is how the CD approach works.

In practice, solving for gauge potential $A$ suffers from a few difficulties. Firstly, the result to Eqs.~\eqref{eq:gauge_A_16} and \eqref{eq: gauge_A} rely on the spectral property of the instantaneous Hamiltonian, which can be hard to obtain. Secondly, the solution may have an ill-defined limit. One situation is when the energy gap closes, such as in evolutions with phase transition. Another case is in a chaotic system, matrix element of $A$ between nearby energy eigenstates are exponentially divergent in system size~\cite{kolodrubetz2017geometry}. Lastly, solutions often contain non-local terms that require a large overhead for implementation. Regarding these difficulties, a variational protocol is proposed which seeks for the best gauge potential within a given operator set which is well-defined.

\subsection{Variational formula for counterdiabatic Approach}
The variational counterdiabatic approach is proposed in Ref.~\cite{sels2017minimizing} based on Eq.~\eqref{eq: gauge_A_19}. Instead of solving for gauge potential $A_\lambda$, one prepares an ansatz of CD terms with tunable parameters. The goal is to find optimal parameters that minimize transitions between eigenstates. 

We define the operator $G_\lambda$ as
\begin{align}
G_\lambda = \partial_\lambda \mathcal{H} + \frac{i}{\hbar}[A_\lambda^*, \mathcal{H}] .
\end{align}
As is clear from Eq.~\eqref{eq: gauge_A_19}, if $A_\lambda^*$ is exact, $G_\lambda = -F_{\text{ad}}$, with no off-diagonal elements. Moreover, the diagonal element of $G_\lambda$ does not depend on $A_\lambda^*$, as the commutator vanishes in this case. Thus, the optimal solution that suppresses transitions among eigenstates is equivalent to reducing the off-diagonal elements of $G_\lambda$ by reducing the Hilbert-Schmidt norm between $G_\lambda$ and $F_{\text{ad}}$, i.e.
\begin{align}\label{eq:action}
D(A_\lambda^*) = \Tr\left[(G_\lambda + F_{\text{ad}})^2\right] = \mathcal{S}(A_\lambda^*) - \Tr[F_{\text{ad}}^2] \hspace{20pt} \mathcal{S}(A_\lambda^*) = \Tr[G_\lambda^2] .
\end{align}
The second equation makes use of the diagonality of $G_\lambda$ and $F_{\text{ad}}$. Since $\Tr[F_{\text{ad}}]$ has no $A_\lambda^*$ dependence, the variational CD approach works by minimizing action $\mathcal{S}(A_\lambda^*)$. Note, that the formula seeks to suppress the average transition rate among all eigenstates. Although it would be better to suppress excitation from the ground state to the first excited state, there is no easy solution.

\subsection{Nested commutator approach}
\subsubsection{Principle of nested commutator approach} 
The nested commutator (NC) formula was proposed in Ref.~\cite{claeys2019floquet} as an efficient and controlled approximation. Even though this ansatz was not used in the main text, we reserve a detailed discussion about it in the Supplementary Material for its popularity, usefulness as well and common confusion. 

\noindent The nested commutator formula reads
\begin{align}\label{supp-eq:10}
\mathcal{A}_\lambda^{(l)} = \ i\sum_{k = 1}^{l} \alpha_k \underbrace{ [\mathcal{H}, [\mathcal{H}, \ldots, [\mathcal{H}}_{2k-1}, \partial_\lambda\mathcal{H}]]] ,
\end{align}
while the matrix element of this expression can be written as 
\begin{align}
\langle m|\mathcal{A}_\lambda |n\rangle = i\sum_{k = 1}^l \alpha_k(E_n- E_m)^{(2k-1)}\langle m|\partial_\lambda\mathcal{H} |m\rangle .
\end{align}
Compare this result to Eq.~\eqref{eq:gauge-compare}, one can see that the variational optimization can then be understood as approximating the exact prefactor $1/(E_n - E_m)$ by a power series $\sum_{k = 1}^l \alpha_{k} (E_n - E_m)^{2k-1}$.  As discussed in Ref.~\cite{claeys2019floquet}, one does not need to worry about the limit $(E_n - E_m)\to 0$ as the relevant energy scale of interests is around energy gap, nor $(E_n- E_m)\gg 1$ as matrix element of $\partial_\lambda\mathcal{H}$ drops in magnitude fast. Thus, it is possible to use finite orders of the formula to approximate exact gauge potential. However, one needs to be careful about the number of orders to keep. Written in the form of summation over different orders, it is tempting to assume that the first order has a dominant effect and that higher orders contribute less. This is not the case. In the following subsection, we will calculate the solution with $l = 1$ and $l=2$ and compare the results.

\subsubsection{First-order approximated gauge potential}
The simplest solution to the nested commutator protocol is the first-order approximation which reads
\begin{eqnarray}\label{supp-eq:11}
\mathcal{A}_\lambda^1 =&& 2\alpha  h_x\big(\sum_i J_i \sigma_y^i + \sum_{i<j} J_{ij} (\sigma_y^i \sigma_z^j + \sigma_z^i \sigma_y^j) + \sum_{i<j<k} J_{ijk} (\sigma_y^i \sigma_z^j \sigma_z^k + 
\sigma_z^i \sigma_y^j \sigma_z^k
+\sigma_z^i \sigma_z^j \sigma_y^k) \nonumber\\
&&+ \sum_{i<j<k<l} J_{ijkl} (\sigma_y^i \sigma_z^j \sigma_z^k\sigma_z^l + 
\sigma_z^i \sigma_y^j \sigma_z^k\sigma_z^l + \sigma_z^i \sigma_z^j \sigma_y^k \sigma_z^l + \sigma_z^i \sigma_z^j \sigma_z^k \sigma_y^l )\big) .
\end{eqnarray}
Here, coefficients $\alpha$ can be solved analytically by minimizing the corresponding action $S$. Plugging the form of $\mathcal{A}_\lambda^1$ in to Eq.~\eqref{eq:action}, and making use of the property that $\Tr(G(\mathcal{A}_\lambda^1)^2$ has non-trivial contributions only when each Pauli operator in $G(\mathcal{A}_\lambda^1)$ multiplies by itself, one obtain the following expression for $S$,

\begin{eqnarray}\label{supp-eq: 23}
S^0(\mathcal{A}_\lambda^1) = \sum_{m=1}^4 (1 + 4(1-\lambda)h^2\alpha m)^2 C_m + h^2\sum_i (1 +  4\lambda \alpha T_{i,i})^2 
+ (8\lambda h\alpha)^2 \sum_{m =2} D_m + (8(1-\lambda)h^2\alpha)^2 \sum_{m = 2}^4 C_m
\end{eqnarray}

with
\begin{align}\label{supp-eq:24}
C_m = \sum_{i_1<\ldots i_m} J_{i1, \ldots <j_m}^2 ,
\end{align}
\begin{align}
T_{j_1\ldots j_p, k_1\ldots k_p} = J_{j1\ldots j_p} J_{k_1\ldots k_q} + \sum_m J_{j1\ldots j_p m} J_{k_1\ldots k_q m}
 + \sum_{m<n} J_{j1\ldots j_p mn} J_{k_1\ldots k_q mn} + \ldots ,
\end{align}
\begin{align}
D_m = \sum_{i_1} \sum_{i_2<\ldots<i_m} S_{i_1, i_2\ldots i_m}^2 , \\
S_{i_1, i_2\ldots i_m} = \sum_{\substack{ j\in S_1\,\, k \in S_2\\ S_1\cap S_2 = \{i_1\}\\S_1\cup S_2 = \{i_1, \ldots i_m\}}} T_{j_1\ldots j_p, k_1\ldots k_q} .
\end{align}
Here, $C_m$ is a sum of square of $m$-local coefficients. For example, $C_1 = \sum_i J_i^2$, $C_2 = \sum_{i<j} J_{ij}^2$, etc.  T is a sum of two-$J$ product determined by its subscript. $T_{j_1\ldots j_p, k_1\ldots k_p} = J_{j_1\ldots j_p} J_{k_1\ldots k_p} + \sum_m J_{j_1\dots j_pl}J_{k_1\dots k_ql} + \sum_{l<m}J_{j_1\dots j_plm}J_{k_1\dots k_qlm}\ldots$  e.g. $T_{12, 34} = J_{12}J_{34} + J_{125}J_{345} + \ldots$ which truncates by interaction order and total qubits.  $S_{i_1, i_2\ldots i_m}$ is a summation over all partitions, such that both $S_1$ and $S_2$ has element $i_1$ with the union contains all $i$'s. For example $S_{1, 234} =T_{1,1234} + T_{12, 134} + T_{13, 124} + T_{14, 123}$.  D is a sum over $S^2$ with a fixed number of indexes in $S$. The sum over $m$ for this term is truncated via the definition of $T$. When $m =1$, $D_1 = \sum_i S_{i,}^2 = \sum_i T_{i,i}^2$. One trick that is helpful is $\sum_i T_{ii} = \sum_i (J_{i}^2 + \sum_j J_{ij}^2 + \sum_{j<k}J_{ijk}^2 + \sum_{j<k<l} J_{ijkl}^2) = \sum_m m C_m$. 

Despite their complex forms, these variables are simply problem-dependent constants and can be easily calculated by programming. Plug Eq.~\eqref{supp-eq: 23} to Eq.~\eqref{eq:action}, the optimal parameter $\alpha$ can be obtained as

\begin{eqnarray}\label{supp-eq:28}
\alpha = -\frac{1}{4}\times
\frac{\sum_{m= 1}^4 m C_m}{(1-\lambda)^2h^2 \sum_{m =1}^4 m^2 C_m + \lambda^2 D_1 + 4\lambda^2 \sum_{m=2} D_m + 4(1-\lambda)^2 h^2 \sum_{m=2}C_m} .
\end{eqnarray}
For 2-local Hamiltonian with $J_{ijk} = J_{ijkl} = 0$, the D terms can be simplified as 
\begin{align}
D_1 = \sum_i T_{i,i}^2 = \sum_i (J_i^2 + \sum_j J_{ij}^2)^2 
= \sum_i J_i^4 + 2\sum_{ij} J_{ij}^4 + \sum_{ij}2 J_i^2 J_{ij}^2 + \sum_{i}\sum_{j<k} 2J_{ij}^2J_{ik}^2 ,
\end{align}
\begin{align}
D_2 = \sum_{i,j} S_{i,j}^2 = \sum_{ij}(T_{i, ij})^2 = \sum_{ij}(J_i J_{ij})^2 , \\
\end{align}
\begin{align}
D_3 = \sum_{i,jk} S_{i,jk}^2 = \sum_{i}\sum_{j<k} T_{ij, ik}^2 = \sum_i\sum_{j<k} J_{ij}^2J_{ik}^2 ,
\end{align}
with $D_m = C_m = 0$ for $m>2$, Equation \eqref{supp-eq:28} can be simplified as 
\begin{align*}
\alpha = -\frac{1}{4}\frac{\sum_i J_i^2 + 2 J_{ij}^2}{R} ,
\end{align*}
\begin{align}
R = (1-\lambda)^2h^2(\sum_i J_i^2+ 8 \sum_{ij} J_{ij}^2) + \lambda^2 (\sum_i J_i^4 + 2\lambda^2\sum_{ij}J_{ij}^4 + 6\sum_{ij}J_i^2 J_{ij}^2 + 6\sum_{i}\sum_{j<k} J_{ij}^2J_{ik}^2) .
\end{align}
This is in accordance to the result in Ref.~\cite{hegade2022digitized}.

\subsubsection{Second order in the nested commutator formula}
The gauge potential corresponding to the second order is more involved. Here, we present the result using symbolic notation
\begin{eqnarray}\label{eq:1st-order}
\mathcal{A}_\lambda^2=&& \sum_i\alpha_y^i \sigma_y^i + \sum_{ij} \alpha_{yz}^{ij} \sigma_y^i \sigma_z^j  + \sum_{ij} \alpha_{yx}^{ij} \sigma_y^i \sigma_x^j + \sum_{ijk} \alpha_{xyz}^{ijk} \sigma_x^i\sigma_y^j \sigma_z^k + \sum_{ijk}\alpha_{yzz}^{ijk}\sigma_y^i \sigma_z^j\sigma_z^k \nonumber\\
&&+ \sum_{ijk}\alpha_{yyy}^{ijk}\sigma_y^i \sigma_y^j\sigma_y^k  + \sum_{ijkl}\alpha_{xyzz}^{ijkl}\sigma_x^i\sigma_y^j \sigma_z^k\sigma_z^l + \sum_{ijkl}\alpha_{yzzz}^{ijkl}\sigma_y^i\sigma_z^j \sigma_z^k\sigma_z^l + \sum_{ijkl}\alpha_{yyyz}^{ijkl}\sigma_y^i\sigma_y^j\sigma_y^k\sigma_z^l .
\end{eqnarray}
Here, $\alpha$'s are coefficients returned by the code, whose explicit form are omitted for brevity. As can be seen from Eq.~\eqref{eq:1st-order}, there are more types of CD terms appears in the second order formula compared to the first order, say XY-type, XYZ-type,YYY-type, XYZZ-type, YYYZ-type. 

Aside from the types, we also compared strength of coefficients with same type of CD term. Consider the term of $\sigma_y^{(0)}$ in the nested commutator with $k=1$ and $k=2$ as an example. For the first order, from Eq.~\eqref{supp-eq:11}, it's easy to see that the coefficient of $\sigma_y^{0}$ is 
\begin{align}
C(\mathcal{A}^{1},\sigma_y^{(0)}) = 2 h_x J_i .
\end{align}
The calculation of the second order is a bit laborious but straightforward. By working out explicitly all Pauli terms corresponding to $k=2$, one can obtain the corresponding coefficient as
\begin{eqnarray}
C(\mathcal{A}^{(1)},\sigma_y^{(0)}) = &&8\lambda^2 h_x\big( J_i ^ 3 + 3 J_i J_{ij}^2 + 3 J_i J_{ijk}^2 + 4 J_{i} J_{ijkl}^2 + 4 J_{ij} J_{ik} J_{ijk}+ 10 J_{ij} J_{ikl} J_{ijkl} \nonumber\\
&&+2 J_{ijk} J_{ijl} J_{ikl} + 8 J_{ijk} J_{ijlm} J_{iklm}\big) + 8 (1-\lambda)^2 h_x^3 J_i .
\end{eqnarray}
Here, the Einstein summation convention is used such that $i$ is a free index that is fixed and all other indices are dummy indices to be summed over. The definition of $J_{ijk}$ is extended such that when $j<i<k$, $J_{ijk} = J_{jik}$. For the case of a 5-qubit problem, that means an expansion from 1 to 212 terms. When the system size $n$ is large, specifically $n>6k$, the number of terms contributing to a single Pauli operator expands as $O(n^6l)$. Moreover, the relation among coefficients becomes more complex, compared to the case of first-order approximation.

\begin{figure}
\centering
\includegraphics[width = 0.7\textwidth]{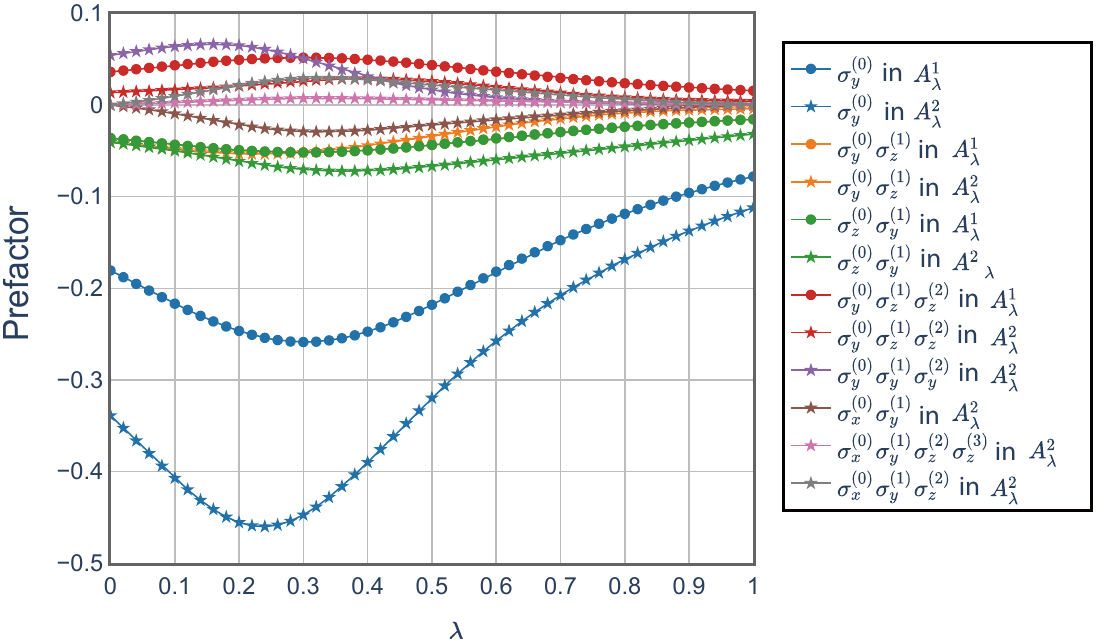}
\caption{$\sigma_y^{(0)}$, $\sigma_y^{(0)}\sigma_z^{(1)}$, $\sigma_y^{(0)}\sigma_z^{(1)}\sigma_z^{(2)}$,  $\sigma_y^{(0)}\sigma_y^{(1)}\sigma_y^{(2)}$ and  $\sigma_x^{(0)}\sigma_y^{(1)}$, $\sigma_x^{(0)}\sigma_y^{(1)}\sigma_z^{(2)}\sigma_z^{(3)}$, $\sigma_x^{(0)}\sigma_y^{(1)}\sigma_z^{(2)}$ for $\mathcal{A}_\lambda^1$ and $\mathcal{A}_\lambda^2$ for $H^{5q}$ and $h_x = 1$. Starred lines represent terms from 1st-order commutator $\mathcal{A}_\lambda^1$ and dashed lines come from 2nd-order nested commutator $\mathcal{A}_\lambda^2$. Coefficients for $\sigma_y^{(0)}\sigma_z^{(1)}$ and $\sigma_z^{(0)}\sigma_y^{(1)}$ are the same for $A_{\lambda}^0$, making the orange star marks overlap with the green star ones. }
\label{supp_fig2}
\end{figure}

To make the discussion more concrete, we consider the $\mathcal{A}_\lambda^{(0)}$ and $\mathcal{A}_\lambda^{(1)}$ for the 5-qubit Hamiltonian described as ~\eqref{eq:h5q} with $h_x = 1$. Here,
\begin{align}
\mathcal{A}_\lambda^2 = \beta_1 [\mathcal{H}, \partial_\lambda \mathcal{H}] + \beta_2 [\mathcal{H}, [\mathcal{H}, [\mathcal{H}, \partial_\lambda \mathcal{H}]]] .
\end{align}
With the help of a symbolic computational program, one can obtain prefactors of various Paul terms in $\mathcal{A}_\lambda^1$ and $\mathcal{A}_\lambda^2$. Here, for $\mathcal{A}_\lambda^{(0)}$, the prefactor of $\sigma_y^{(0)}$ is obtained as $\alpha C_{\mathcal{A}^{1}}(\sigma_y^{(0)})$ and for $\mathcal{A}_\lambda^{(1)}$, the corresponding prefactor is calculated as $\beta_1 C_{\mathcal{A}^{1}}(\sigma_y^{(0)}) + \beta_2 C_{\mathcal{A}^{2}}(\sigma_y^{(0)})$, where parameters $\alpha$ and $\beta$'s are the optimial solution that minimizes the action $S$.  Figure \ref{supp_fig2} shows a subset of these prefactors.  As is clear from the graph, the dominant contribution for this case comes from 1-local terms like $\sigma_y^{(0)}$, whose optimal solution shows substantial discrepancy between $\mathcal{A}_\lambda^1$ and $\mathcal{A}_\lambda^2$. Moreover, the 2nd order result also comes with a richer set of operators which manifest themselves away from the initial and final points

As discussed in Ref.~\cite{passarelli2020counterdiabatic}, a general case of $l\ge 3$ is needed for a system of size $n<100$ with two-local or three-local interactions. Moreover, the order $l$ needed to suppress excitation is shown to grow with the number of qubits and interaction order. In Ref.~\cite{hatomura2021controlling}, YX$\dots$XZ type of operators up to full length are considered. It was observed that including long-range terms are necessary to suppress transitions as evolution time increases.

\subsection{User Defined Ansatz}
Although NC formula provides a well-controlled approximation to the CD ansatz, the result often involves highly non-local terms. As discussed in Ref.~\cite{passarelli2020counterdiabatic}, a truncated version with a selected subset of local CD terms can produce comparable results to a high order of NC results, say 3-local CD terms can compete with $l=8$ CD solution. The form of user-defined ansatz has a dominant effect on the efficacy of the approach. In this work, we explored 5 user-defined CD ansatz with up to 2-local interactions. These user-defined ansatz are

\begin{figure}
\centering
\includegraphics[width=0.7\textwidth]{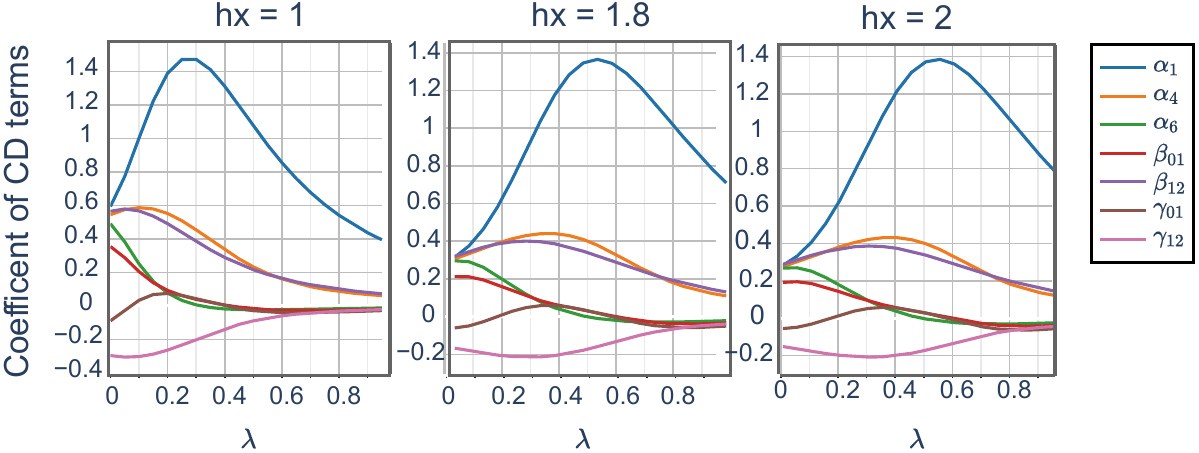}
\caption{A subset of solutions of coefficients of the (Y+YZ+ZY)-type CD ansatz for $h_x = 1$, 1.8 and 2 respectively. The problem considered is $H^{9q}$.}
\label{supp_fig3}
\end{figure}

\begin{enumerate}
\item Y-type \hspace{92pt} $\mathcal{A} = \sum_i \alpha_i J_i\sigma_y^i $
\item (Y + YZ${}_{\text{u}}$)-type \hspace{53pt}$\mathcal{A} = \sum_i \alpha_i J_i \sigma_y^i + \beta \sum_{i<j} J_{ij}\sigma_y^i\sigma_z^j $
\item (Y + YZ${}_{\text{u}}$ + ZY${}_{\text{u}}$)-type \hspace{20 pt}$\mathcal{A} = \sum_i \alpha_i J_i \sigma_y^i + \beta \sum_{i<j}J_{ij}\sigma_y^i\sigma_z^j  +\gamma \sum_{i<j} J_{ij}\sigma_z^i\sigma_y^j$
\item (Y + YZ)-type \hspace{53pt} $\mathcal{A} = \sum_i \alpha_i J_i \sigma_y^i + \sum_{i<j} \beta_{ij}J_{ij}\sigma_y^i\sigma_z^j $
\item (Y + YZ + ZY)-type \hspace{23pt} $\mathcal{A} = \sum_i \alpha_i J_i \sigma_y^i + \sum_{i<j} \beta_{ij}J_{ij}\sigma_y^i\sigma_z^j + \sum_{i<j} \gamma_{ij} J_{ij} \sigma_z^i\sigma_y^j $
\end{enumerate} 
Here, $\alpha$s, $\beta$s and $\gamma$s are variational parameters to be fixed by Eq.~\eqref{eq:action}. Figure \ref{supp_fig3} displays a subset of solutions of these coefficients for different $h_x$. As is shown in the figure, each solution is a bell-shaped function that peaks at different $\lambda$s for different $h_x$. Experimenting with other $h_x$'s, it is observed that with increasing $h_x$, these peaks shift to the right monotonically. This explains the result in Figure 1 of the main text that performance varies for different $h_x$. A boost in accuracy occurs when the sampled points in a trotterized evolution coincide with a peak in the solution. Figure 1 also reveals an unexpected fact that $p = 1$ in most cases has the best performance in terms of accuracy and the gain is amplified for evolutions with CD contribution only. It is not entirely clear why this is the case, but it definitely is good news for the NISQ era which says that the counterdiabatic approach is able to produce decent results with the simplest version of it.
 
\begin{figure}
\centering
\includegraphics[width=0.75\textwidth]{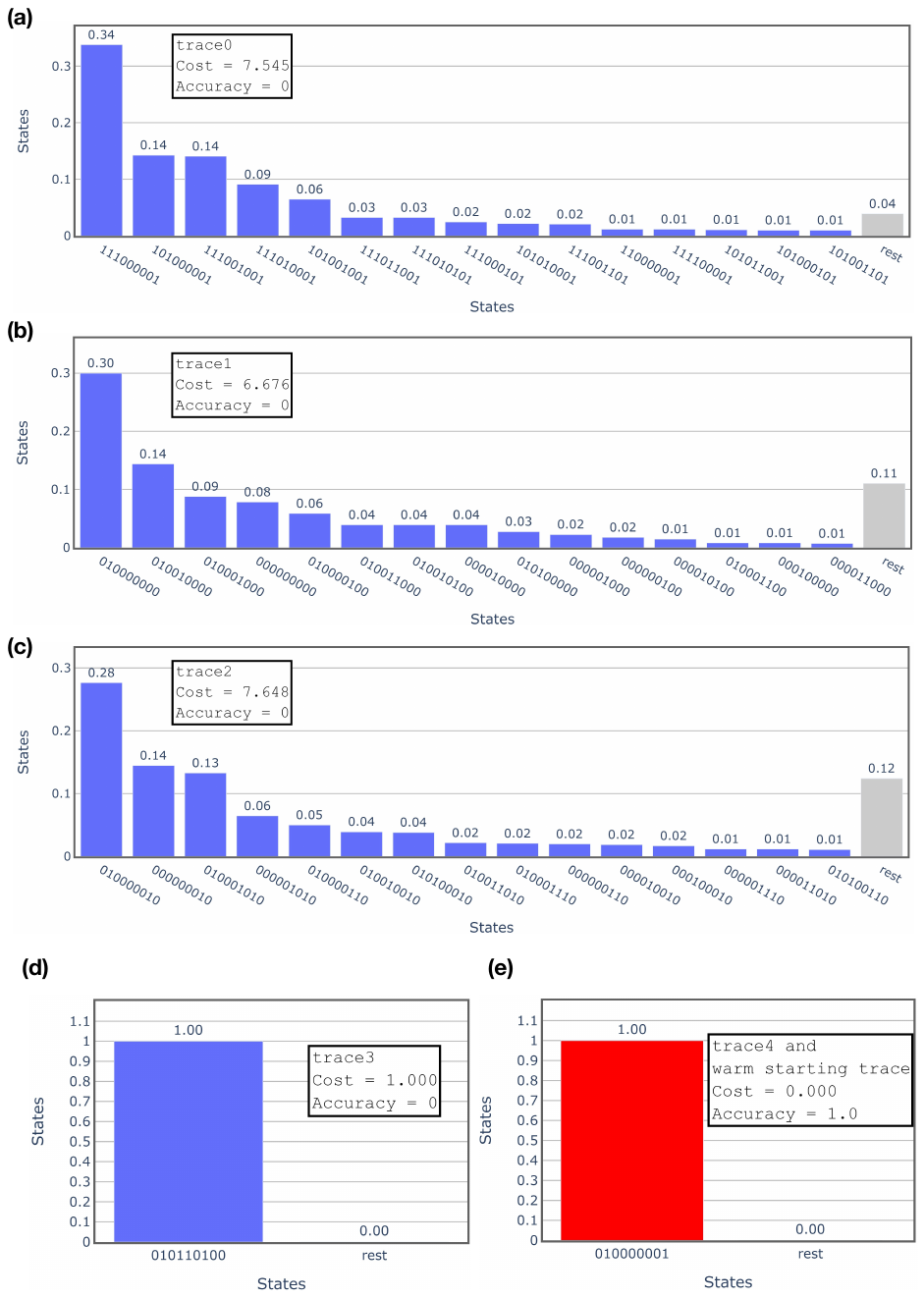}
\caption{Final states distribution after performing a single layer of CD approach with Y, Y+YZ\_u, Y+YZ\_u+ZY\_u, Y+YZ, Y+YZ+ZY type of CD terms respectively. The left y axis and bar chart is for probability distribution and right y axis and scattered plot is for cost. Blue bars correspond to the most likely 15 states and red bar is to the target state, grey bar describes the total probability for unaccounted states. Average cost and accuracy are listed in the annotation. Target Hamiltonian is $H^{9q}$.}
\label{supp_fig4}
\end{figure}

In Fig.~\ref{supp_fig4}, we compare the performance of the five user-defined CD ansatz for $p = 1$ for solving $H^{9q}$. Parameters are fixed by minimization of action and $h_x$ is tuned to maximize sum of coefficients at $t = \tau/2$. As seen from the distribution, except for the Y-type ansatz, in all other scenarios, the target state takes up a  considerable amount of the total occurrence with (Y+YZ+ZY)-type having the best performance. For the latter, the target state is the most probable state and the corresponding accuracy for this ansatz is $12\%$. All other ansatz generates an accuracy of $8\% - 10\%$, which provides a large amount of simplifications in the analytical analysis and implementation without greatly sacrificing the performance. To see this, note that the optimal solution is obtained by solving a system of linear equations whose number equals that of the parameters, which are$ n$ $n+1$, $n+2$, $(n^2 + n)/2$, $n$ for ansatz of type 1 to 5 respectively. Moreover, the number of coupling terms also varies greatly, which amount to 0, $n(n-1)/2$, $n(n-1)$, $n(n-1)/2$, $n(n-1)$ for each of them, leading to a circuit depth which scales up with the increased number. Moreover, the system has one state with energy 1 which corresponds to the lowest excitation. This state appears with a higher probability in ansatz 5 than in ansatz 2. This makes the 2nd ansatz both efficient and more favorable for further optimization as it is less likely to be trapped in this suboptimal point.

\subsection{Warm Starting}
Aside from using the results obtained from the variational approach as it is, one can also use it as a warm start based on which further optimization is made to improve the performance and concentrate the target state. In this section, we will adopt the ($\textrm{Y} + \textrm{YZ}_{\textrm{u}}$)-type of CD ansatz and explore the parameter space of $\alpha$s and $\beta$s for $H^{9q}$. Parameters will be optimized with ADAM optimizer so as to minimize the average energy of the final state. Default options are used for the optimizer with step size set to be 0.01, $\beta_1$ and $\beta_2$ set as 0.9 and 0.99 respectively. 

\begin{figure}
\centering
\includegraphics[width = 0.9\textwidth]{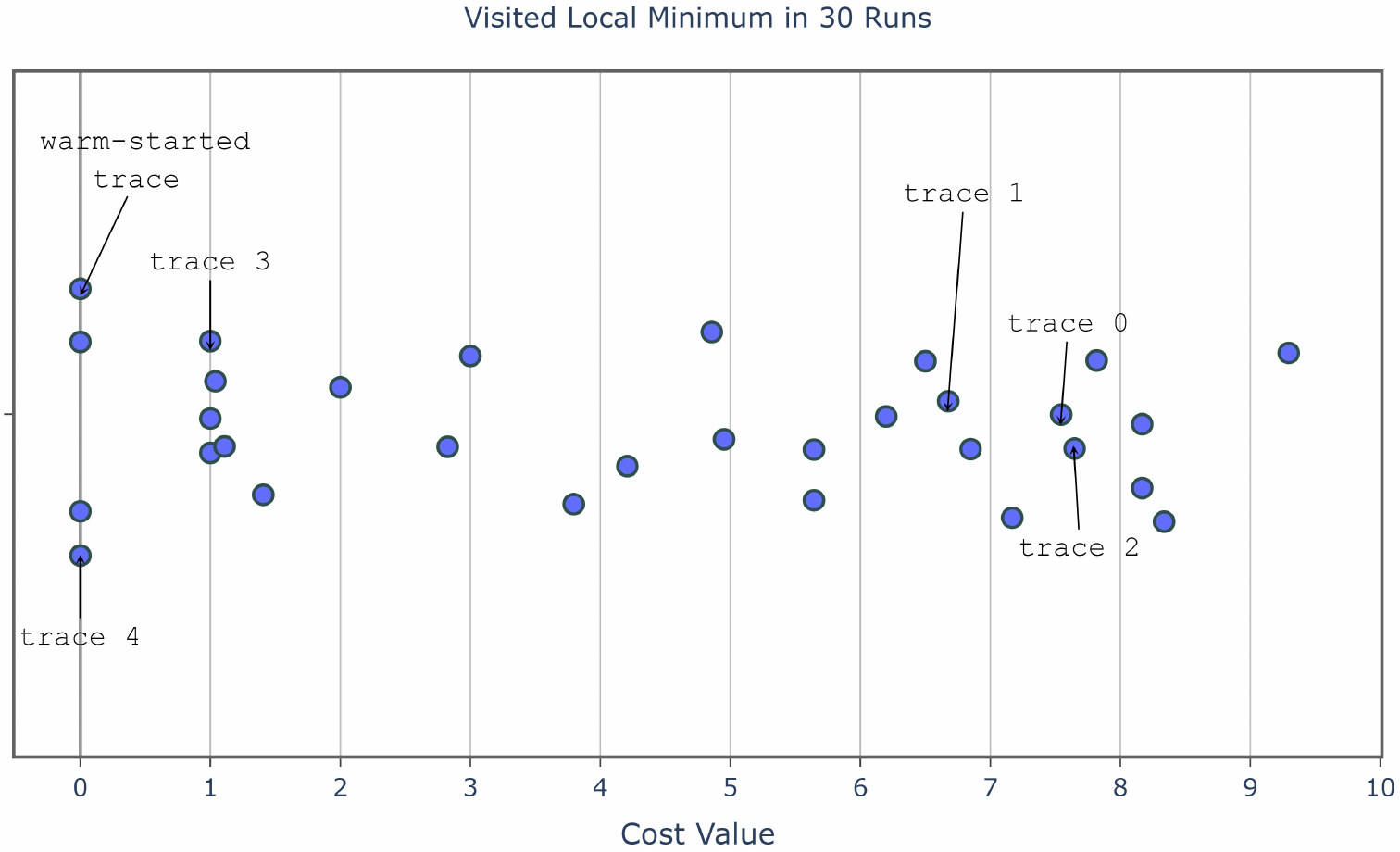}
\caption{Distribution of trapped local minimum in 30 runs of optimization with random initialization. The x axis represents a cost. The y axis is for jittering the date for better visualization. Traces labels are in accordance with those in Figure~2 of the main text. The three points lying on the zero cost function indicate the true ground state. All other points represent local minimums}
\label{supp_fig5}
\end{figure}

Figure~2 in the main text displays the evolution of the cost function for optimization from CD solutions as well as five processes with random initialization. As we can see from that plot, the warm-started optimization successfully converges to the true ground state with cost being zero. Four of the five randomly initialized optimizations get struck at local minimum. These 6 processes are further examined in the space of cost energy and parameter space $\alpha_1$ and $\beta$, see Fig.~\ref{supp_fig5}. Note, the parameter space is a reduction from the full 10-dimensional ones where only changes in the two retained dimensions are plotted, all other 8 parameters are updated according to the gradients during the optimization process. As is consistent with Fig.~2 in the main text, warm-started trace and trace 4 converge to the same points in the parameter space that corresponds to the desired solution. Other randomly-initialized traces converge to a variety of local minimums. Moreover, the only successful trace(trace 4) starts with a higher cost value compared with others. This implies that an initial point with low-cost values does not guarantee a good starting point for optimization and vice versa. 

\begin{figure}
\includegraphics[width=0.85\textwidth]{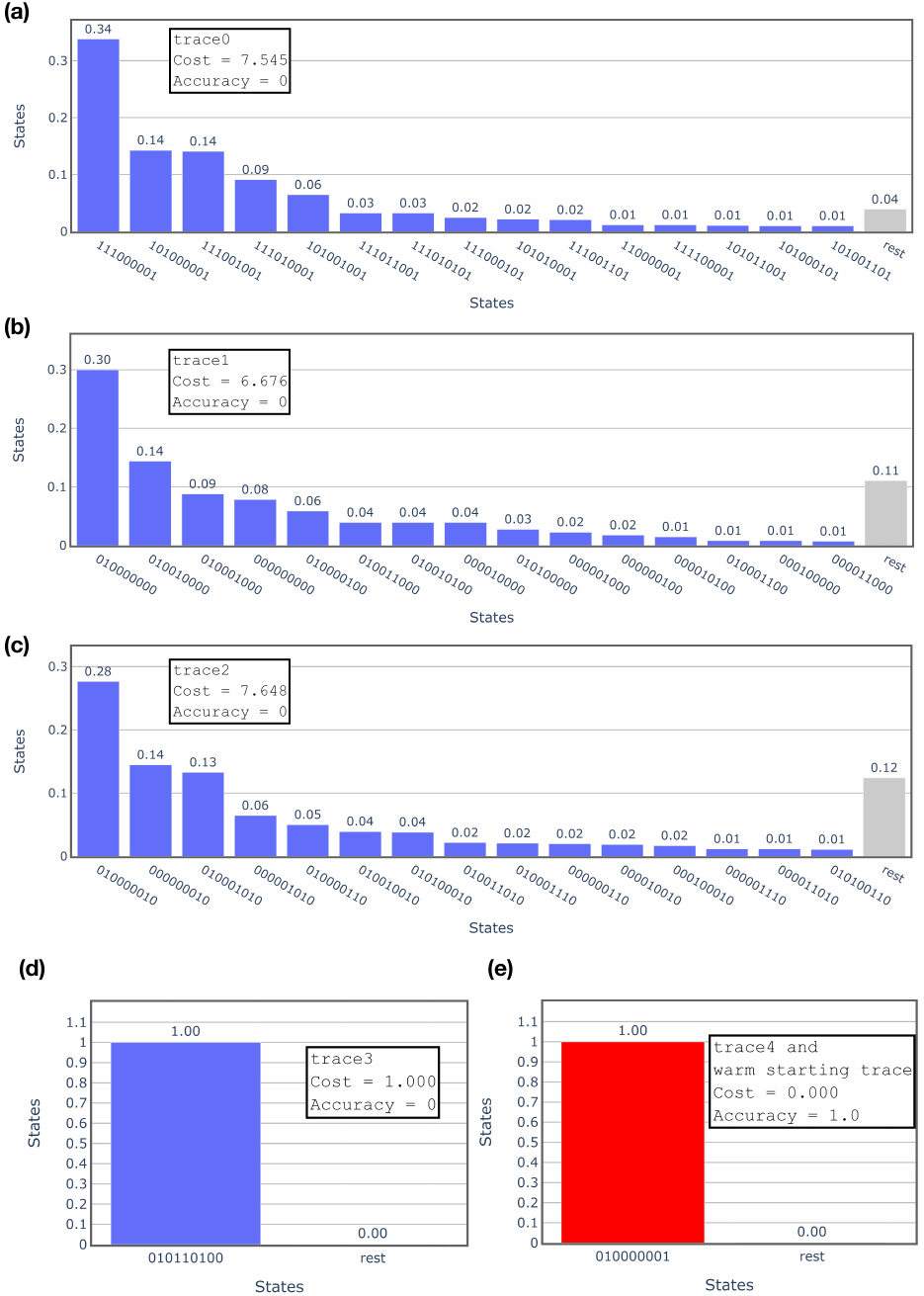}
\caption{State distributions generated by evolving the system with p = 1 (Y+YZ)-type of CDs ansatz. Parameters are taken at converging points in the optimization processes labeled by trace 0 to trace 4 as well as warm starting trace. Trace labels are consistent with the ones used in Figure 4 in the main text}
\label{supp_fig6}
\end{figure}

We repeated the randomly-initialized optimization 30 times and at the end of each iteration, Hessian matrices are calculated to obtain the index of the trapped point. In all 30 cases, the number of negative eigenvalues of the Hessian is 0, indicating that all of the points are local minimum. It is found that 3 out of 30 runs converge to global minimum point while others converge to diversified local minimums, see Fig.~\ref{supp_fig5} where local minimum has a wide span in the cost value. Among the listed 5 random traces, trace 4 converges to the ground state, and trace 3 converges to the first excited eigenstate. The other 3 traces are all trapped in a mixed state, see Fig.~\ref{supp_fig6}. Thus it can be inferred, that the cost energy landscape is very complex, where the number of local minimum is more than the number of eigenstates in the system. Thus, a warm starting point that stays near the ground state is of great value to an optimization process in such a system.

\section{Algorithm Implementation}\label{sec: transpile}
\subsection{Exponentiation of Operators}
\begin{figure}
\includegraphics[width=0.7\linewidth]{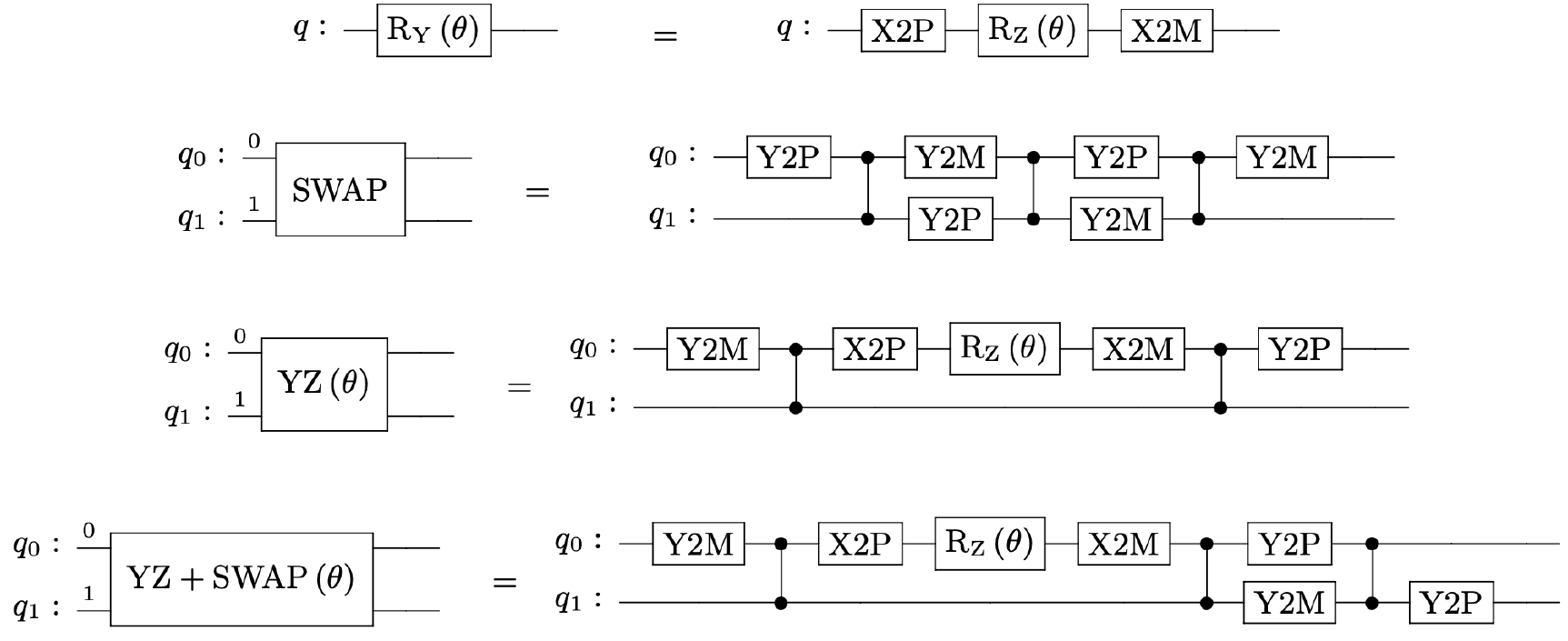}
\caption{Gate composition of $\text{R}_{\text{Y}}$, SWAP, YZ and YZ+SWAP gate using native gates of quantum processor}
\label{supp_fig7}
\end{figure}

The implementation of CD approach can be reduced to exponentiating Y-type and $YZ$-type operators. This can be realized easily using the design proposed in \cite{weidenfeller2022scaling}. In this method, $\exp(-iJ\sigma_i^y)$ can be implemented via $\text{RY}_i(2\theta)$ gate and $\exp(-i J\sigma_i^y\sigma_j^z)$ can be realized as $\text{CX}_{ij}\text{RY}_i(2J)\text{CX}_{ij}$. Figure~\ref{supp_fig7} shows the construction of basic building blocks of the algorithm in terms of native gates of the hardware. Here X2P(Y2P), X2M(Y2M) are rotations around X(Y) axis by degree $\pi/2$ and $-\pi/2$ respectively. $R_Z(\theta)$ is rotations around Z axis by angle $\theta$. Lines with endpoints denote CZ gates. In the plot, we also show implementation of SWAP gates and a combination of YZ and SWAP. The latter combination can reduce the CZ gates by two, thus is preferred whenever possible.

\subsection{Swap strategy}

Aside from efficient ways to exponentiate the operators, another practical issue with implementation is the swap strategy. Due to limited connectivity, swap gates are needed to implement gates that are not compatible with the hardware structure. Different topologies, as well as swapping strategies, have a large influence on the number of swap gates and the circuit depth of the algorithm. In this work, we adopt the method proposed in \cite{weidenfeller2022scaling}. 

\begin{figure}
\centering
\includegraphics[width = 0.85\textwidth]{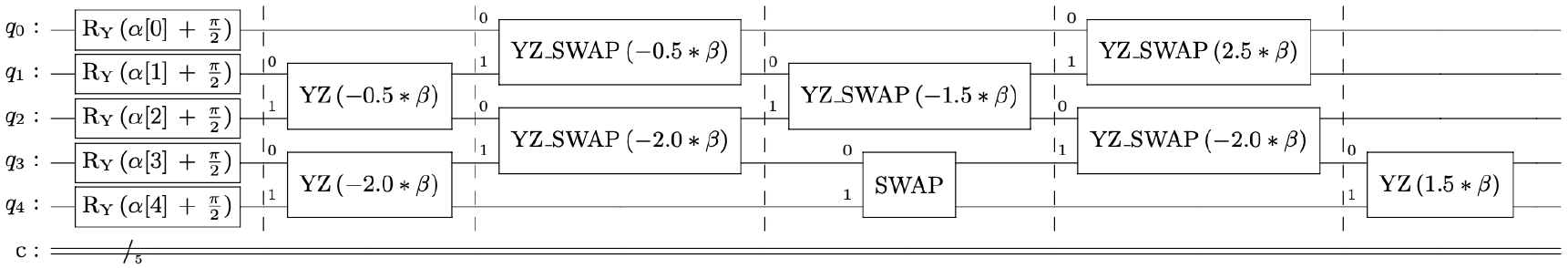}
\includegraphics[width = 0.85\textwidth]{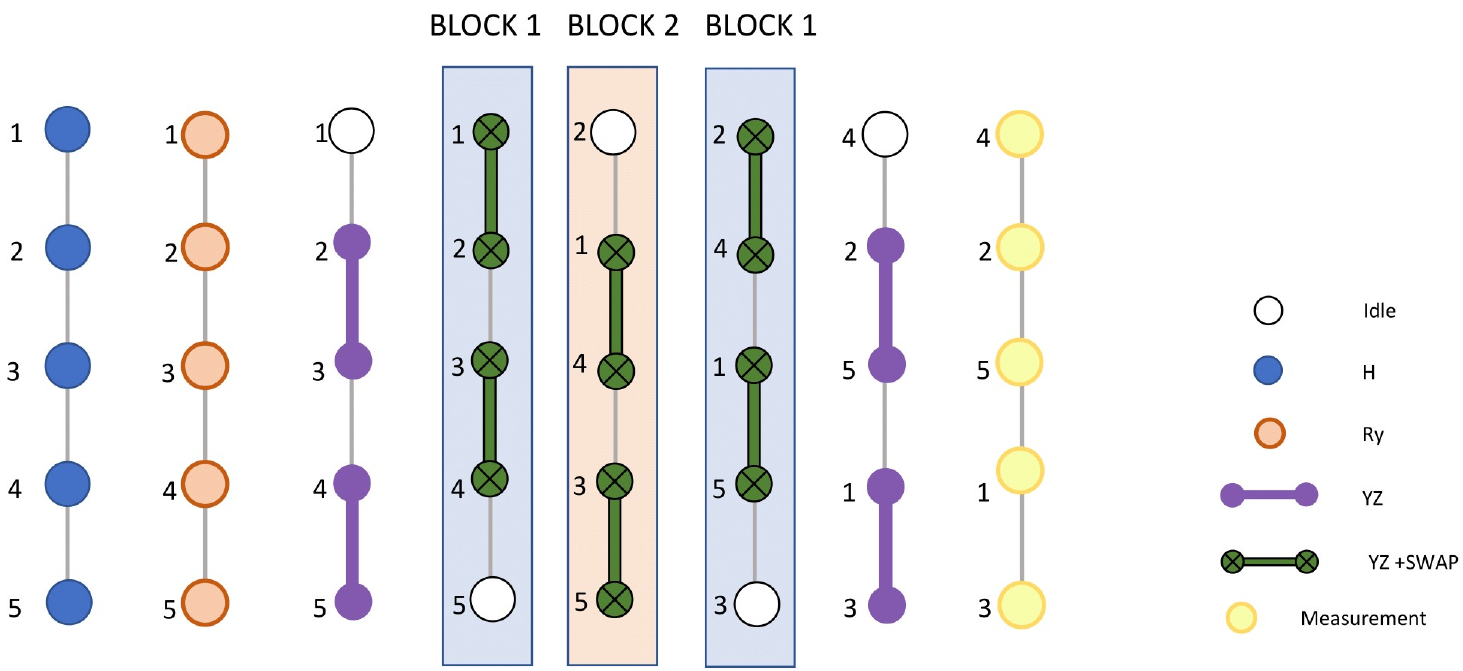}
\caption{Circuit diagram for implementing a 5 qubit swap strategy on a 2 by 6 grid. The upper graph is in the representation of gate components and the lower graph is in the diagram. Two representations are equivalent. Circles and dumbbells representations as gates are listed in the legend. Blocks are categorized by the swap pattern implemented. Numbers on the dots are to track shuffles among qubits.}
\label{supp_fig8}
\end{figure}

\begin{figure}
\includegraphics[width = 0.8\textwidth]{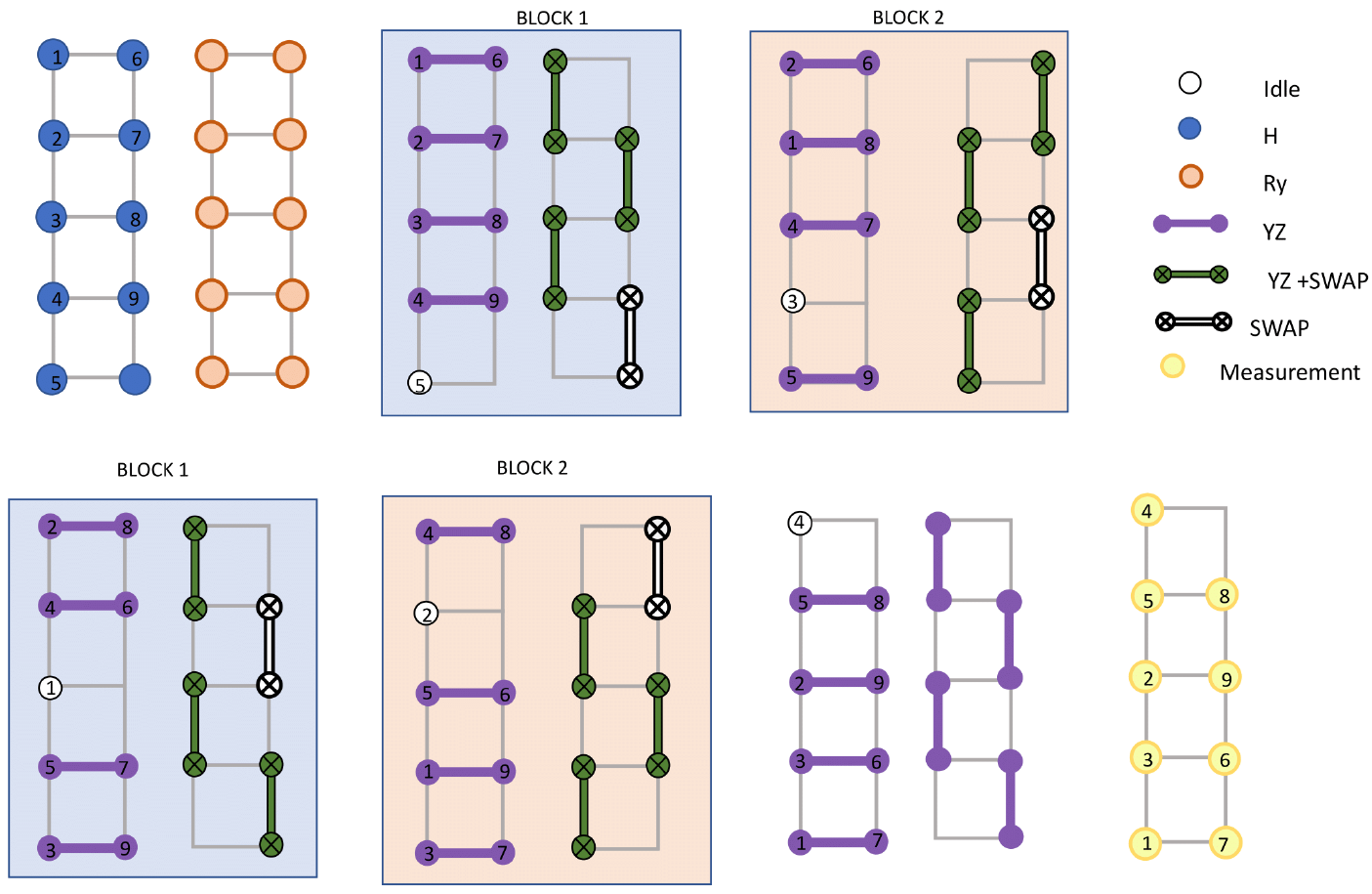}
\label{fig:sub1}
\includegraphics[width = 0.8\textwidth]{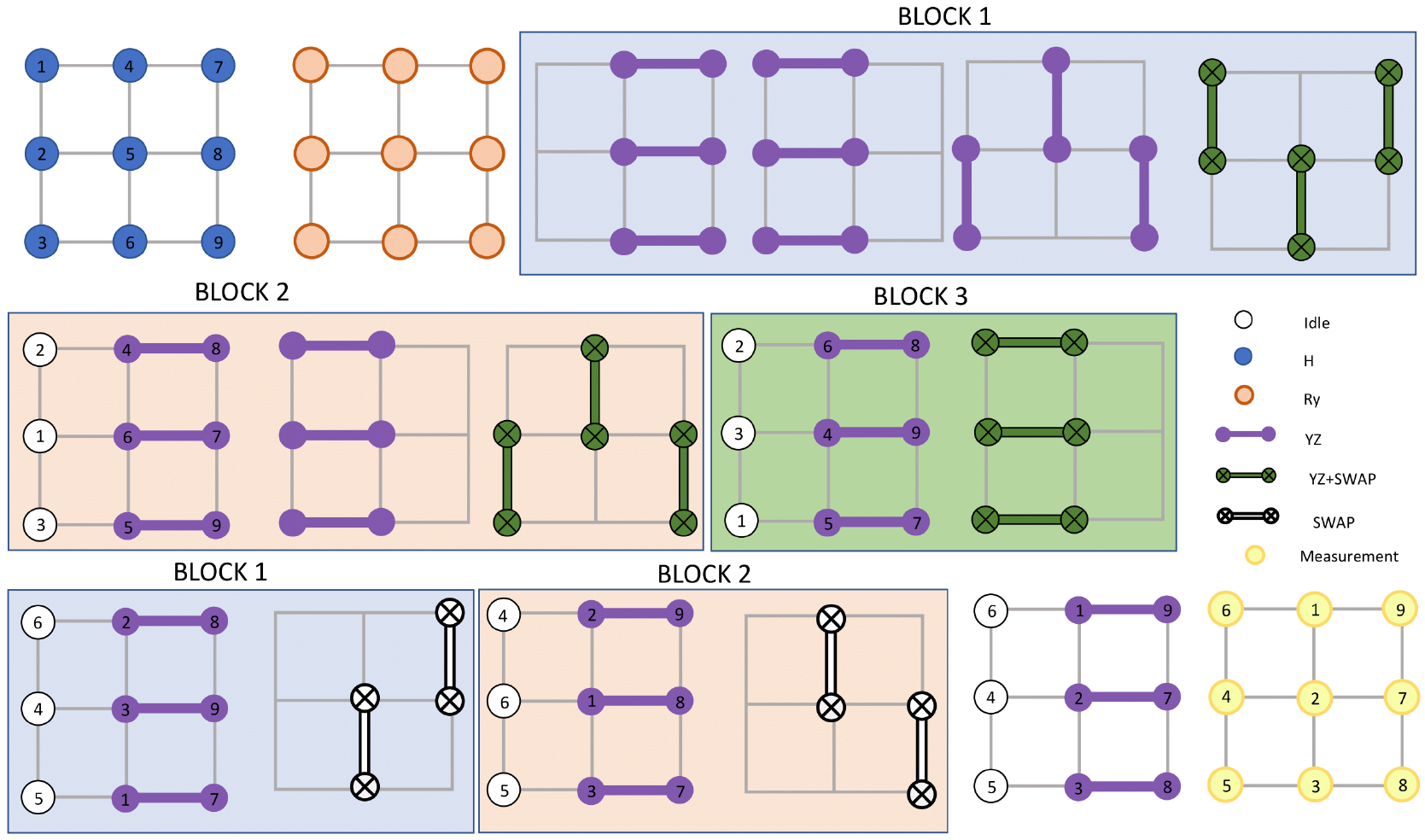}
\label{supp_fig9}
\caption{Color online. Circuit Diagram for implementing a 9-qubit swap strategy. The upper diagram is implemented on a 2 by 5 grid while the lower diagram is on a 3 by 3 grid. In the upper strategy, the lower right corner qubit is ancillary and remains at $|0\rangle$ state. Only swap gates are applied to this qubit which moves it around and enables interactions among other working qubits. Different blocks are for different swapping patterns which are represented by a YZ+SWAP gate(green) or a SWAP gate(white). In the upper graph, there are two types of blocks corresponding to two verticle swapping patterns. The lower graph has three patterns, including a new horizontal swapping.  Numbers are shown to track shuffling among qubits.}
\label{supp-fig:3}
\end{figure}

\begin{figure}
\centering
\includegraphics[width = 0.8\textwidth]{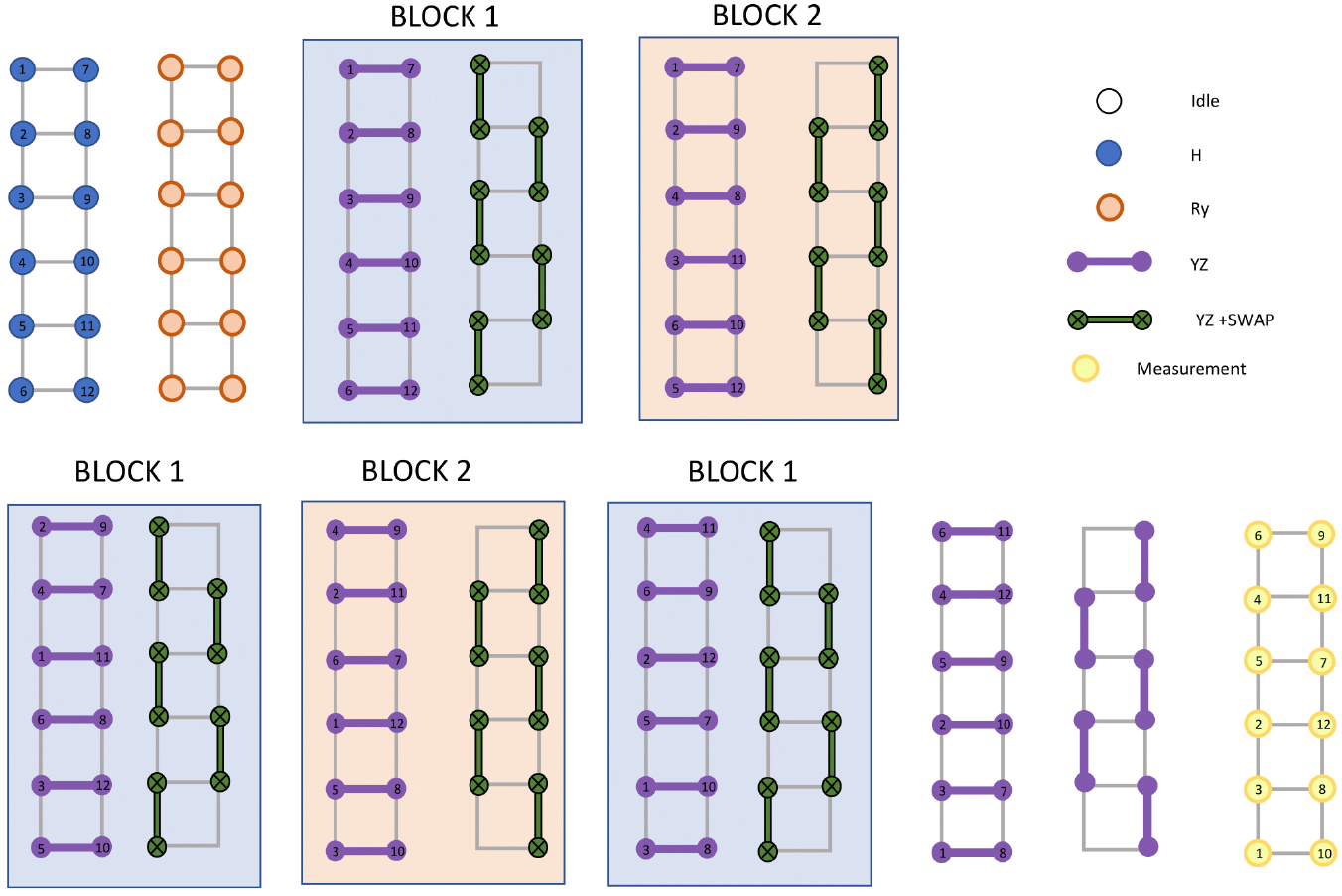}
\caption{Circuit diagram for implementing a 12 qubit swap strategy on a 2 by 6 grid. }
\label{supp_fig9m}
\end{figure}

In this swapping strategy, qubit topology is described by an $n$ by $m$ grid. First, swapping is applied in vertical direction until all qubits in the same line or adjacent line have been connected once($n-1$ swap layers). Then swapping along the horizontal direction is made in steps of two to shuffle columns. See Figs.~\ref{supp_fig8}, \ref{supp_fig9}, \ref{supp_fig9m} for examples of systems with 5, 9, and 12 qubits. In these figures, a set of parallel swap gates as well as YZ interactions compliant with the hardware topology are grouped into blocks, and different blocks are labelled by the swapping pattern. 

\subsection{Circuit depth for {\it n}-qubit system with {\it p}=1}

Based on this swap strategy, the total number of swap blocks, swap gate numbers, layer counts for YZ, YZ+SWAP, and SWAP layer for different topology are calculated, see Table~\ref{supp_tab1},\ref{supp_tab2} and ~\ref{supp_tab3}. As each swap gate comprises 3 CZ gates, reducing the swap gate count would alleviate the effect of gate errors. Each layer comes with a fixed duration, reducing the layer numbers would lower the circuit duration, thus lessening the decoherence effect.

\begin{table}
  \label{supp_tab1}
  \includegraphics[width=0.7\linewidth]{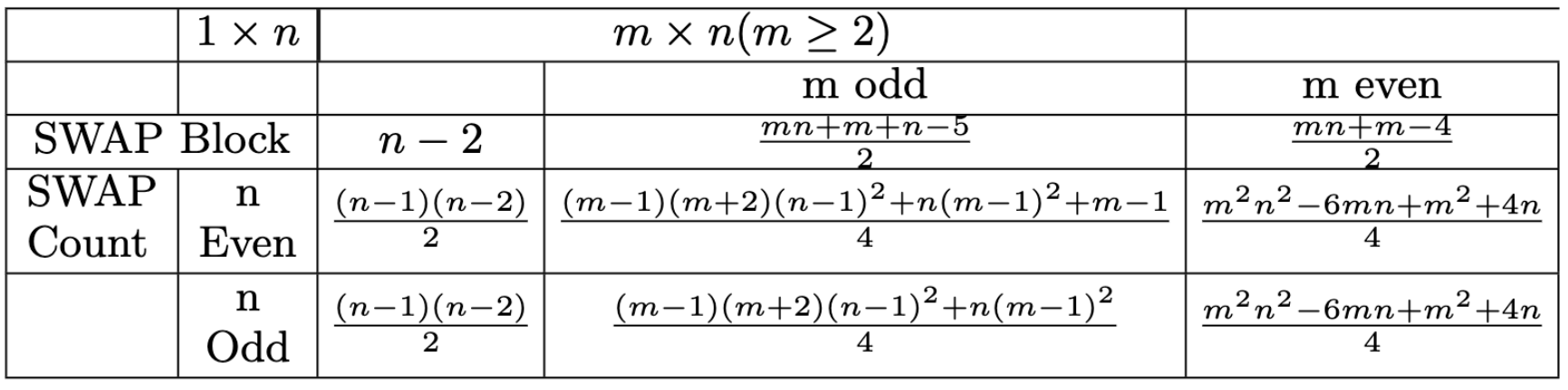}
  \caption{SWAP block count and SWAP gate Count for Various Topology.}
\end{table}


\begin{table}
  \label{supp_tab2}
  \includegraphics[width=0.7\linewidth]{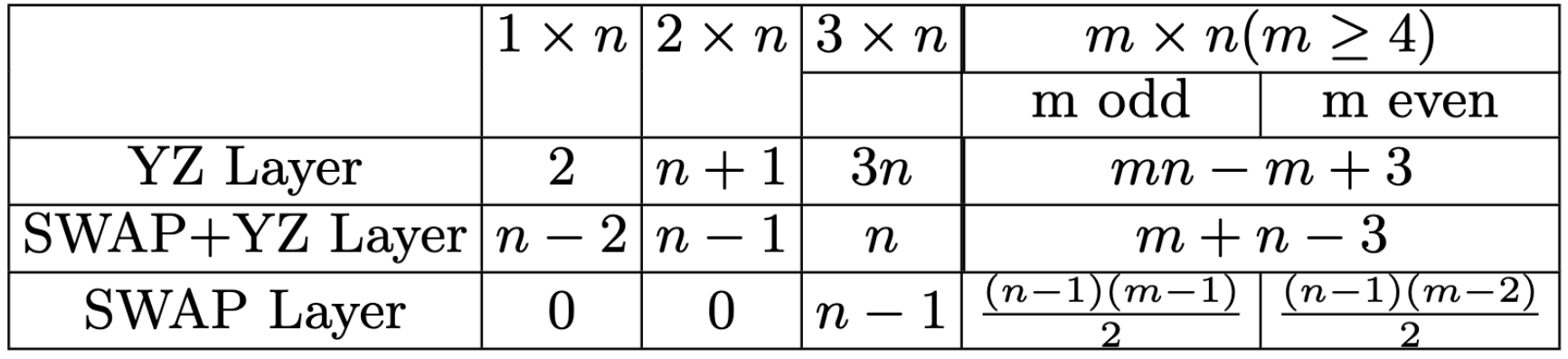}
  \caption{Layer Count for Various Topology for Y + YZ or Y + YZ${}_{\text{u}}$ type of CD ansatz.}
\end{table}

\begin{table}
  \label{supp_tab3}
  \includegraphics[width=0.7\linewidth]{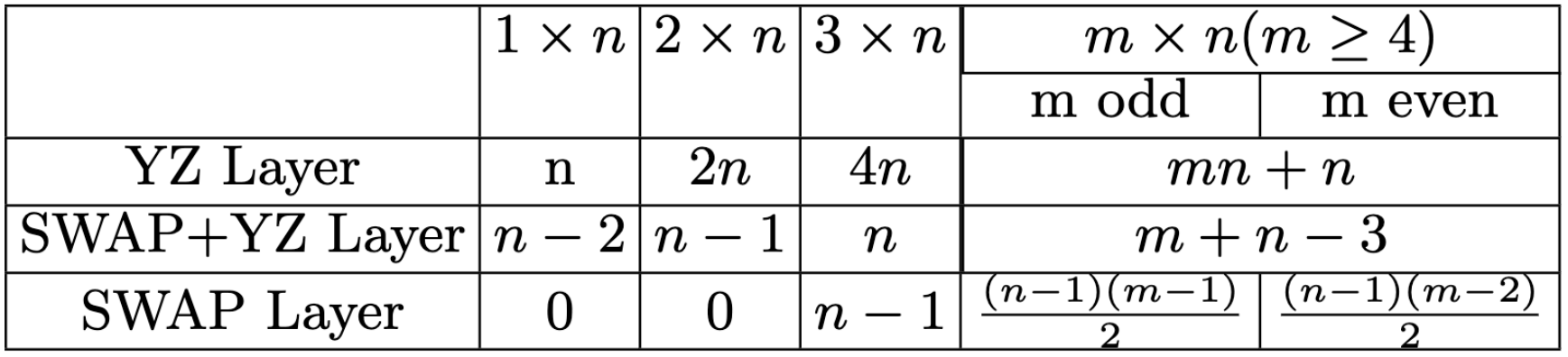}
  \caption{Layer Count for Various Topology for Y + YZ + ZY or Y + YZ${}_{\text{u}}$+ ZY${}_{\text{u}}$ type of CD ansatz.}
\end{table}

\begin{figure}
\centering
\includegraphics[width = 0.9\textwidth]{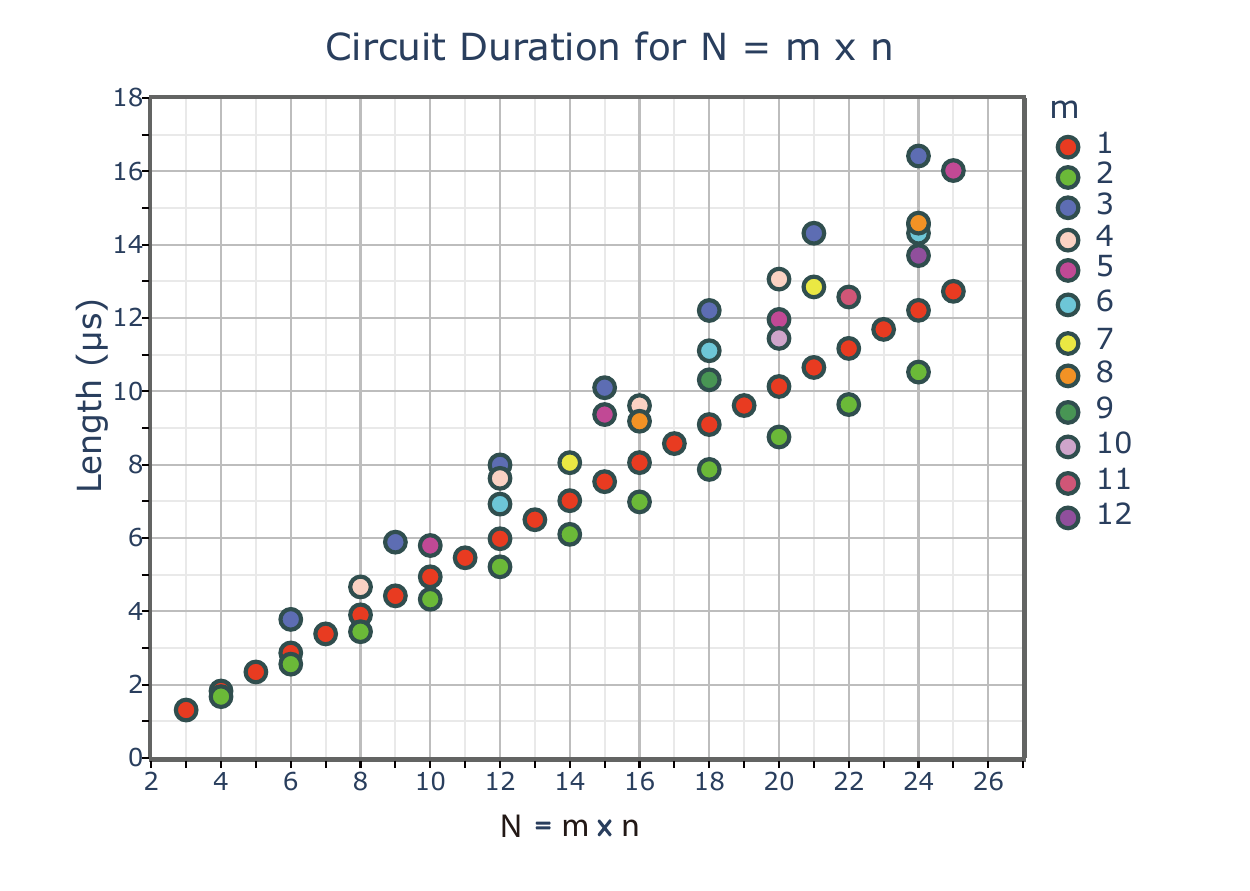}
\caption{Circuit duration for various topologies. The x-axis is for the total qubit number. The y-axis describes circuit duration. Colors are related to the column number of the grid.}
\label{supp_fig10}
\end{figure}

Based on the information in table \ref{supp_tab2}, we also plot the circuit duration of the CD approach with (Y + YZ)- and (Y + YZ${}_{\text{u}}$)- type of ansatz. The duration for single qubit gates and two-qubit gates are set as 30ns and 123ns according to the parameters of the hardware. The result is plotted in Fig.~\ref{supp_fig10}. As is clear from the plot, the optimal topology in terms of circuit depth of an $N$-qubit system is a 2 by $N/2$ grid, rather than a square grid.  The shape with 3 columns has the longest circuit duration. Note the column and row index play different roles in the swap strategy, thus $m$ and $n$ are not symmetric. The graph provides guidance for possible system sizes accessible under coherent time constraints. In our system, the T2* is around 5$\mu s$, thus the largest system we have explored is 12-qubit. One can adjust the gate construction scheme and gate duration according to the hardware at hands and plot a similar graph for their own reference. A similar plot can be constructed using \ref{supp_tab3} for CD approach with (Y + YZ+ ZY)- and (Y + YZ${}_{\text{u}}$ + ZY${}_{\text{u}}$)- type of ansatz, which is omitted for brevity.

\section{Device Information}\label{sec:device}
\begin{figure}
\includegraphics[width = 0.75\textwidth]{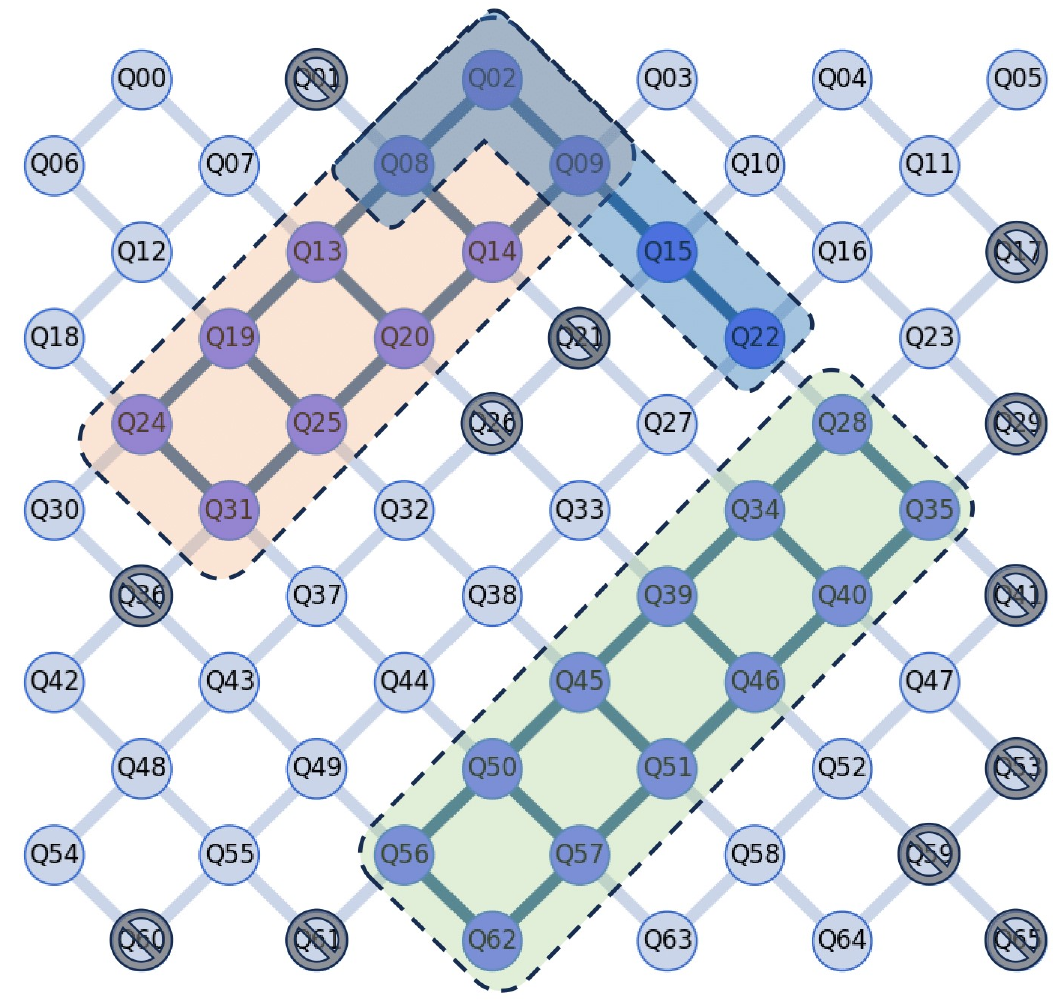}
\caption{Quantum processor layout diagram. Bad qubits are marked by a cross-out sign. Qubits used in the $H^{5q}$, $H^{9q}$, and $H^{12q}$ experiments are highlighted and shaped in blue, orange, and green colors respectively.}
\label{supp_fig11}
\end{figure}

For our experiment, we use the superconducting quantum computer from Quantumctek\cite{QuantumCteck}. The processor is arranged in a 6 by 11 grid as shown in Fig.~\ref{supp_fig11}. It consists of 54 tunable-frequency transmon qubits, with microwave-driven single and two-qubit gates. The native single qubit gates are X2P, X2M, Y2P, Y2M, and Rz which describe rotation around the x-axis by $\pi/2$, $-\pi/2$, rotation around the y-axis by $\pi/2$, $-\pi/2$ and rotation around z axis by an arbitrary angle respectively. The native two-qubit gate is the CZ gate which is realized by parametric-resonance interaction via coupler. The qubit properties and gate characterization for those used in the experiment for $H^{5q}$, $H^{9q}$ and $H^{12q}$ are listed in Table \ref{supp_tab4}, \ref{supp_tab5} and \ref{supp_tab6}. Single qubit gate duration is 30ns, CZ gate duration is 123ns. 
\begin{table}
  \label{supp_tab4}
  \includegraphics[width=0.7\linewidth]{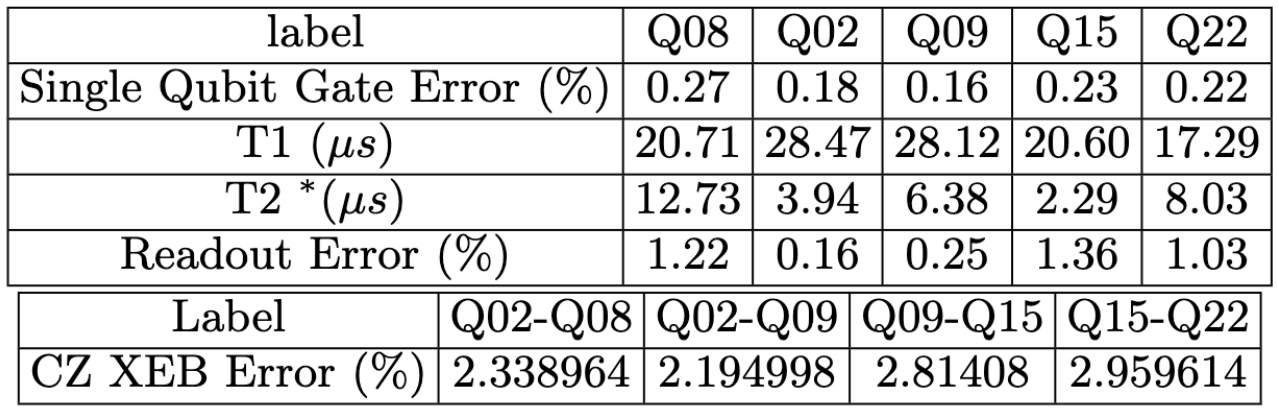}
  \caption{Characterization of qubits used in $H^{5q}$ experiment.}
\end{table}
\begin{table}[!hbt]
\centering
\begin{tabular}{ |*9{c|}}  
\hline
Label &  Q02 &    Q08 &    Q13 &    Q19 &    Q24 &    Q09 &    Q14 &    Q20  \\
\hline
Single Qubit Gate Error (\%)&   0.18 &   0.27 &   0.42 &   0.21 &   0.31 &   0.16 &   0.21 &   0.55  \\
\hline
T1    ($\mu s$)                  &  28.47 &  20.71 &  16.08 &  25.29 &  17.83 &  28.12 &  26.83 &  20.71  \\
\hline
T2*   ($\mu s$)                &   3.94 &  12.73 &   3.58 &   1.58 &   2.47 &   6.38 &   4.46 &   7.39  \\
\cline{1-2}
Readout Error (\%)         &   0.16 &   1.22 &   0.75 &   0.88 &   1.71 Label &   0.25 &   1.93 &   2.58 \\
\hline
{} &   Q25 &    Q31 \\
\cline{1-3}
Single Qubit Gate Error &  0.33 &   0.15 \\
\cline{1-3}
T1                      &  9.71 &  25.65 \\
\cline{1-3}
T2*                     &  3.24 &  11.17 \\
\cline{1-3}
Readout Error           &  0.25 &   1.21 \\
\cline{1-3}
\end{tabular}

\vspace*{0.5cm}

\begin{tabular}{|*9{c|}}
\hline
Label &   Q02-Q08 &   Q02-Q09 &   Q08-Q13 &   Q08-Q14 &   Q09-Q14 &   Q13-Q19\\
\hline
CZ XEB Error (\%) & 2.34 &     2.19 &     2.95 &     4.11 &     4.12 &     2.36 \\
\hline
Label  &   Q13-Q20  &  Q14-Q20 &   Q19-Q24 &  Q19-Q25 &   Q20-Q25 &   Q24-Q31  \\
\hline
CZ XEB Error (\%) &  2.35 &     2.88 &     2.82 &     2.77 &     2.57 &      2.9  \\
\hline
Label  &   Q25-Q31 \\
\cline{1-2}
CZ XEB Error (\%) &  4.01\\
\cline{1-2}
\end{tabular}
\caption{Characterization of qubits used in $H^{9q}$ experiment.}
\label{supp_tab5}
\end{table}

\begin{table}[!hbt]

\centering
\begin{tabular}{|*9{c|}}
\hline
Label &    Q28 &    Q34 &    Q39 &    Q45 &    Q50 &    Q56 &    Q35 &    Q40  \\
\hline
Single Qubit Gate Error (\%) &   0.21 &   0.22 &   0.16 &   0.28 &   0.19 &   0.15 &   0.44 &   0.23   \\
\hline
T1                      &  25.77 &  20.39 &  28.92 &  24.85 &  22.98 &  28.75 &  18.93 &  29.20  \\
\hline
T2*                    &   6.66 &   1.48 &   2.11 &   1.88 &   0.77 &   2.08 &   7.57 &   8.11  \\
\hline
Readout Error  (\%)         &   0.27 &   0.56 &   1.53 &   0.19 &   1.20 &   0.61 &   0.19 &   0.57 \\
\hline
Label & Q46 &    Q51  &    Q57 &    Q62 \\
\cline{1-5}
Single Qubit Gate Error (\%) &  0.59 &   0.21 &     0.21 &   0.27 \\
\cline{1-5}
T1       ($\mu s$)       &  7.58 &  27.70          &   26.49 &  17.78 \\
\cline{1-5}
T2*       ($\mu s$)       &  7.27 &   2.90       &     2.61 &   5.18 \\
\cline{1-5}
Readout Error  (\%)   &  0.46 &   1.44        &     0.29 &  22.81 \\
\cline{1-5}
\end{tabular}

\vspace*{0.5cm}

\begin{tabular}{|*7{c|}}
\hline
Label &  Q28-Q34 &  Q28-Q35 &  Q34-Q39 &  Q34-Q40 &  Q35-Q40 &  Q39-Q45 \\
\hline
CZ XEB Error (\%)&     2.24 &     2.38 &     2.92 &     1.94 &     3.12 &     2.69  \\
\hline
Label  &  Q39-Q46 &  Q40-Q46 &  Q45-Q50 &  Q45-Q51 &  Q46-Q51 &  Q50-Q56 \\
\hline
CZ XEB Error (\%) &     3.96 &     2.75 &      1.5 &     2.44 &     3.51 &     2.97   \\
\hline
Label  &  Q50-Q57 &  Q51-Q57 &  Q56-Q62 & Q57-Q62\\
\cline{1-5}
CZ XEB Error (\%) &     2.85 &     2.83 &     3.34 &      3.2 \\
\cline{1-5}
\end{tabular}
\caption{Characterization of qubits used in $H^{12q}$ experiment.}
\label{supp_tab6}
\end{table}
\end{document}